\documentclass[aps, amsmath, amssymb, nofootinbib, citeautoscript, twocolumn, floatfix, superscriptaddress]{revtex4}

\usepackage{graphicx}          
\usepackage{dcolumn}           
\usepackage{multirow}          
\usepackage{bm}                
\usepackage{amsmath}           
\usepackage{verbatim}          
\usepackage{xcolor}            
\usepackage{subfigure}         
\usepackage{wrapfig}           
\usepackage{mathrsfs}          
\usepackage[version=4]{mhchem} 
\usepackage[symbol]{footmisc}  
\usepackage[utf8]{inputenc}    
\usepackage{natbib}            

\usepackage[colorlinks,citecolor=blue,linkcolor=blue,urlcolor=blue,bookmarks=false,hypertexnames=true]{hyperref}
\usepackage{orcidlink}         

\begin{document}


\title{Zonal Flow Excitation in Electron-Scale Tokamak Turbulence}

\author{Stefan Tirkas \orcidlink{0000-0002-8050-617X}}
\email{stefan.tirkas@colorado.edu}
\affiliation{University of Colorado at Boulder, Boulder, CO 80309, United States of America}
\author{Haotian Chen \orcidlink{0000-0003-3896-4369}}
\affiliation{Southwestern Institute of Physics, Chengdu 610041, China}
\author{Gabriele Merlo \orcidlink{0000-0003-4877-1456}}
\affiliation{The University of Texas at Austin, Austin, TX 78712, United States of America}
\author{Frank Jenko}
\affiliation{Max Planck Institute for Plasma Physics, Boltzmannstr 2, 85748 Garching, Germany}
\author{Scott Parker}
\affiliation{Renewable and Sustainable Energy Institute, University of Colorado at Boulder,\\ Boulder, CO 80303, United States of America}

\date{\today}

\begin{abstract}
   \quad The derivation of an intermediate-scale gyrokinetic-electron theory in nonuniform tokamak plasmas [Chen H. \textit{et al} 2021 \textit{Nucl. Fusion} \textbf{61}
   \href{https://doi.org/10.1088/1741-4326/abf81a}{066017}] has shown that a Navier-Stokes type nonlinearity couples electron-temperature-gradient (ETG) modes and zonal flow
   (ZF) modes with wavelengths much shorter than the ion gyroradius but much longer than the electron gyroradius. This intermediate-scale ETG-ZF coupling is typically stronger
   than the Hasegawa-Mima type nonlinearity characteristic of the fluid approximation and is predicted to lead to relevant zonal flow generation and ETG mode regulation.
   Electron-scale, continuum, gyrokinetic simulation results are presented here which include both single-mode ETG and full-spectrum ETG turbulence. The zonal flow generation
   due to single ETG modes is investigated and the single-mode intermediate-scale results are found to be in agreement with theory. The full-spectrum results are then presented
   and explained qualitatively in terms of the single-mode results. It is found that the ETG-driven zonal flows regulate intermediate-scale electron heat flux transport to levels
   in the predicted range.

   \medskip\noindent\hfil\rule{.5\linewidth}{0.1pt}\hfil
\end{abstract}

\maketitle

\section{\label{sec:Intro}Introduction}

\quad Electron-scale turbulence is a plausible explanation for the anomalous transport of electron energy well above the neoclassical values seen in a variety of tokamak
plasma scenarios \cite{BatchTrans,ConfineITER,HortonTransport,DoyleTransport}. Additionally, electron energy transport may become more important in future burning plasma
experiments such as ITER because the electron channel is preferentially heated by Coulomb collisions with fusion alpha particles. The electron-temperature-gradient (ETG)
instability produces radially-elongated streamers at the electron gyroradius scale and is a primary candidate to explain electron-scale transport \cite{ConfineITER,
DoyleTransport,LeeETG,DorlandJenkoETG1,DorlandJenkoETG2,DorlandJenkoETG3,Nevins06}. Electron heat flux due to ETG turbulence has been seen to play a role in various
tokamak experiments \cite{RenETG,GriersonETG,RyterETG,KieferETG}, and the inclusion of electron-scale dynamics at the ion scale has resulted in better agreement with
experimental heat flux levels \cite{HowardLmode,HollandHmode}.

\quad Electron-scale \cite{ParkerZF,ColyerETG} and multiscale \cite{HowardLmode,HollandHmode} long-time, large-box gyrokinetic flux-tube simulations have reported that
intermediate-scale zonal flows (ZFs) help to regulate streamer turbulence in the quasi-saturated state and can eventually become dominant. These results are inconsistent
with fluid ETG turbulence models in which zonal flow generation occurs via the standard Hasegawa-Mima nonlinear mechanism \cite{HasegawaMima}, which is significantly
weaker for ETG turbulence than for ion-temperature-gradient (ITG) turbulence \cite{LinFluidETG,ChenFluidETG}. Moreover, while shearing due to zonal flows generated by
ITG turbulence can suppress ion transport levels, the finer scale ETG turbulence is unlikely to be affected by the ITG-driven zonal flows \cite{BatchTrans,KimETG}. These
effects have led to the expectation of a streamer-dominated steady state at the electron scale.

\quad A weak-turbulence, toroidal, gyrokinetic-electron analysis \cite{Chen_ETG_ZF} of nonuniform tokamak plasmas in the intermediate-scale ($k_{\perp}^2\rho_e^2\ll 1
\ll k_{\perp}^2\rho_i^2$) results in a Navier-Stokes type nonlinearity which is typically stronger than the Hasegawa-Mima coupling of the fluid approximation. Here
$k_{\perp}$ is the wavenumber perpendicular to the magnetic field and $\rho$ is the particle gyroradius. The subscripts $i$ and $e$ denote ions and electrons respectively.
This stronger Navier-Stokes type nonlinearity is predicted to drive notable ZF generation and ETG mode regulation at intermediate scales when compared to the Hasegawa-Mima
type coupling of the short-wavelength fluid regime \cite{Chen_ETG_ZF}.

\quad ETG mode saturation in the short-wavelength fluid regime has been extensively studied and includes saturation mechanisms such as secondary instabilities \cite{DorlandJenkoETG1,
DorlandJenkoETG2} and toroidal inverse-cascading \cite{LinFluidETG,ChenFluidETG,KimETG} which would lead to a turbulent state characterized by intermediate-scale ETG modes.
After the initial transition to the intermediate scale, the nonlinear interaction between ETG and ZF modes is expected to be enhanced such that zonal flows may grow to regulate
long-term steady state transport levels as measured by experiment \cite{Chen_ETG_ZF}. The intermediate-scale turbulent state might then be characterized by kinetic saturation
mechanisms such as standard quasilinear estimates \cite{JenkoSat1,JenkoSat2,BourdSat} and $\textbf{E}\times\textbf{B}$ particle trapping \cite{ParkerITG}.

   \begin{figure*}[!ht]
      \centering
      \subfigure[]{
      \label{fig:BenchmarkITG}
      \includegraphics[width=2.4in, height=1.9in]{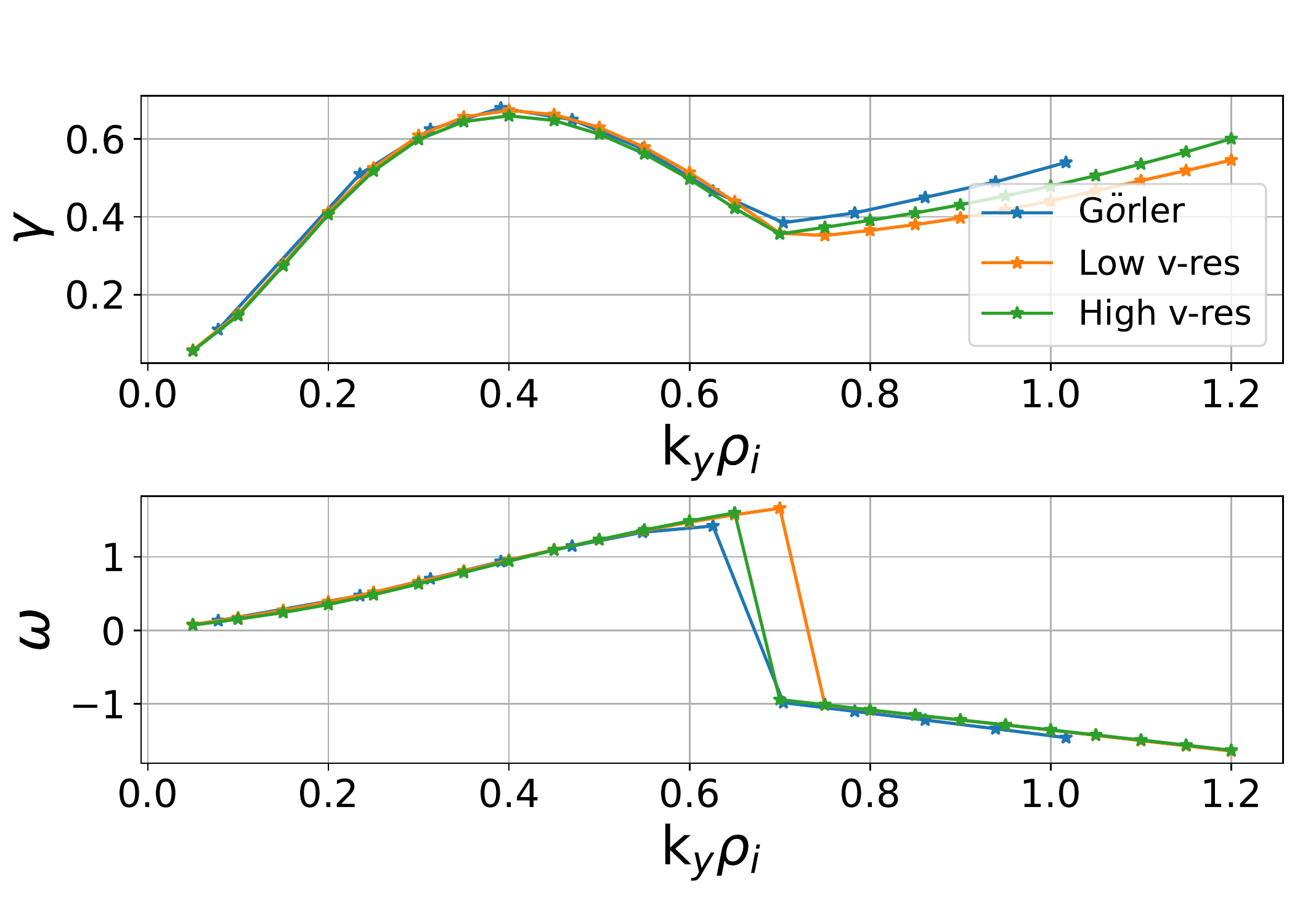}} 
      \subfigure[]{
      \label{fig:GrowthRatesETG}
      \includegraphics[width=2.2in, height=1.9in]{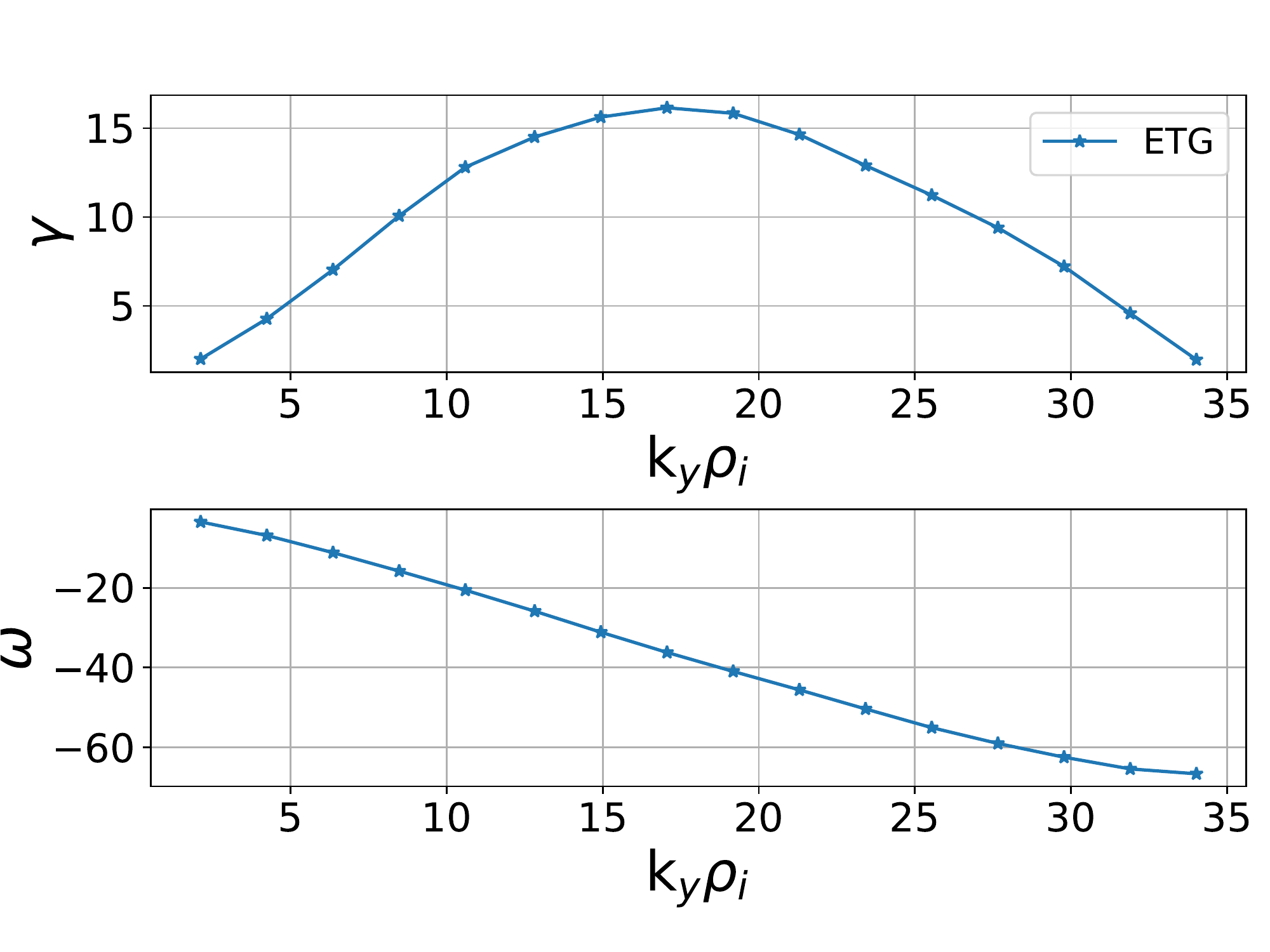}}
      \subfigure[]{
      \label{fig:GrowthRatesTEM}
      \includegraphics[width=2.2in, height=1.9in]{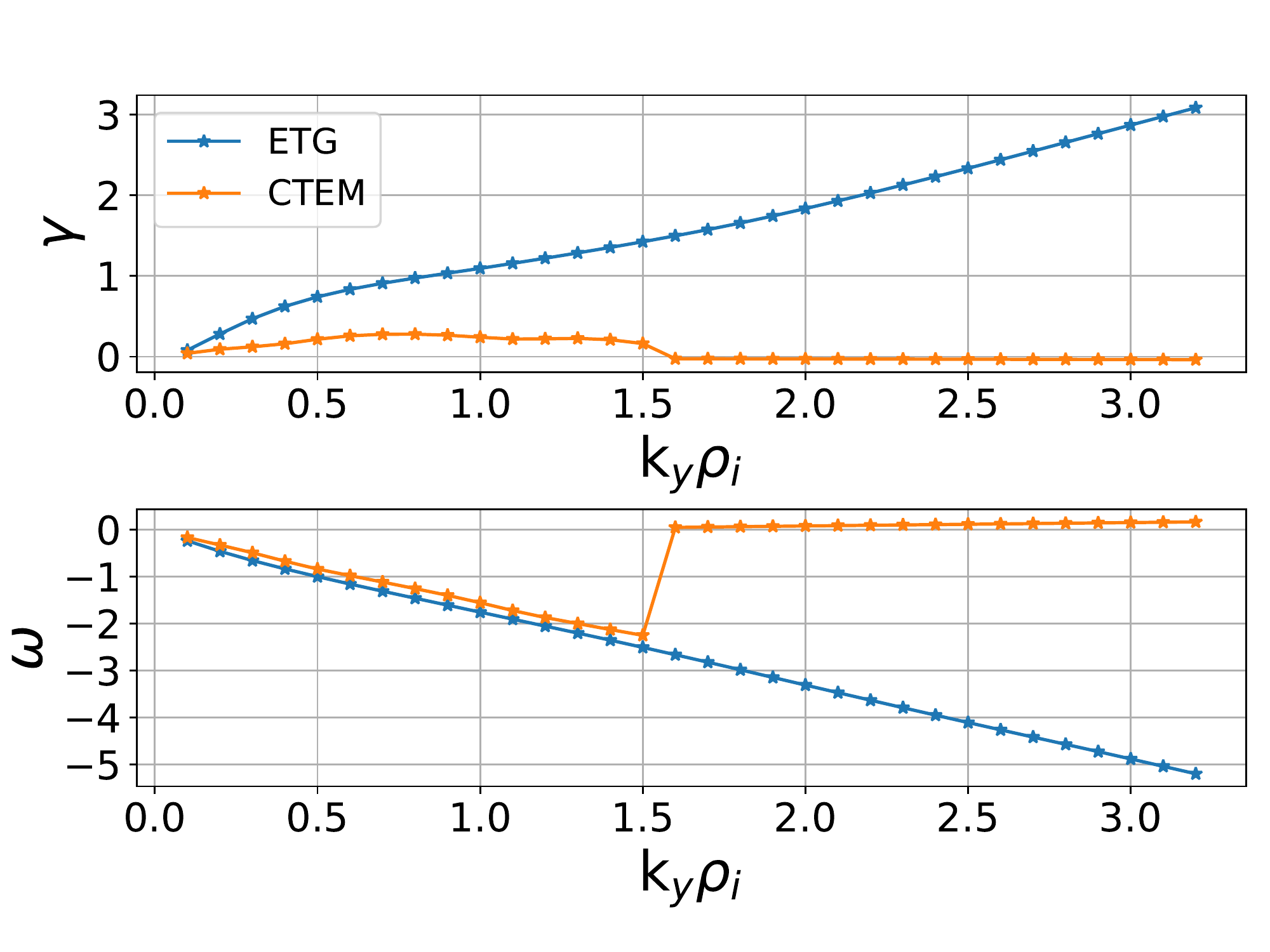}}
      \caption{GENE linear simulation results showing growth rates and real mode frequencies for (a) CBC ITG benchmark case, (b) electron-scale ETG case, and (c) ion-scale ETG
      case showing CTEM modes as well. Positive(negative) frequencies indicate propagation in the ion(electron)-diamagnetic direction.}
      \label{GrowthRates}
   \end{figure*}

\quad Collisionless Cyclone base-case (CBC) simulations of ETG turbulence initially reported the algebraic growth of zonal flows into late times \cite{ParkerZF}.
Electron-scale MAST simulations \cite{ColyerETG} demonstrated that the long-time saturated electron heat flux scales roughly proportionally to the collisionality,
and this was tied to the nonlinear interaction of ETG modes with zonal flows which are well-known to be damped by collisions \cite{DiamondZonal}. In both cases, an
initial turbulent state developed characterized by ETG mode streamers which were eventually suppressed by the slow growth of intermediate-scale zonal flows. DIII-D
and Alcator C-Mod simulations \cite{HollandHmode,HowardLmode} involving multiscale ion and electron dynamics also saw significant intermediate-scale zonal flow
generation which helped to suppress ITG and ETG turbulence into the late stage.

\quad This paper provides results which compare the generation of zonal flow by ETG turbulence in electron-scale, gyrokinetic simulations with theoretical predictions
in the intermediate-scale limit \cite{Chen_ETG_ZF}. We first provide the details of the simulation parameters in Section \hyperref[sec:Model]{II}. This is followed by
analysis of two types of nonlinear simulations. The ``single-mode" results serve to illuminate the role of a single ETG mode in generating zonal flow and are covered in
Section \hyperref[sec:SingleMode]{III}. As the theoretical description is limited to a single ETG mode for tractability \cite{Chen_ETG_ZF}, these results convey the primary
scope of this paper. The ``full-spectrum" nonlinear simulations provided in Section \hyperref[sec:FullSpec]{IV} include a typical range of ETG modes and are qualitatively
explained in terms of the single-mode results. In both types of simulations it is found that intermediate-scale zonal flows are primarily driven by slowly-saturating
intermediate-scale ETG modes. The results are in good agreement with the gyrokinetic theory in the intermediate-scale and the electron fluid models at the short wavelength
scale.

\section{\label{sec:Model}Simulation Model and Parameters}

\quad We employ GENE \cite{DorlandJenkoETG1,GeneCode}, an Eulerian 5-d gyrokinetic continuum code, in the flux-tube limit appropriate for electron-scale turbulence.
Gyrokinetic ions and electrons are taken with standard Cyclone base-case (CBC) parameters which are typical of H-mode core plasmas, but here a simplified circular geometry
is used \cite{CBC_1,CBC_2,CBC_3,GorlerBenchmark}. First, the linear ion-scale benchmark in Ref. \cite{GorlerBenchmark} was verified, then the simulation was converted to
the electron scale by reducing the perpendicular box dimensions by a factor of $\sqrt{m_i/m_e} \sim 42$. Here, $m_e$ is the electron mass, and $m_i$ is the ion mass which
is taken to be the proton mass, $m_p$. For the electron-scale case the ion temperature gradient was set to zero to suppress long-wavelength ion turbulence and focus on
electron-scale physics.

\quad The mode frequencies, $\omega$, and growth rates, $\gamma$, resulting from the linear GENE simulations are shown as functions of $k_y\rho_i$ in Fig. \ref{GrowthRates}
above. The frequencies are normalized to units of $R/c_s$, as listed with other GENE normalizations in Table \hyperref[tableOne]{I}. Here, $k_y$ is the binormal wavenumber
of the GENE coordinate system, $R$ is the tokamak major radius, and $c_s = \sqrt{T_i/m_e}$ is the ion sound speed with $T_i$ the ion temperature. The ITG benchmark case is
shown in Fig. \ref{fig:BenchmarkITG} alongside the electron-scale ETG case in Fig. \ref{fig:GrowthRatesETG}. Ion-scale ETG results are shown in Fig. \ref{fig:GrowthRatesTEM},
where the collisionless trapped-electron mode (CTEM) \cite{CBC_TEM,AdamCTEM,ChenCTEM} is included. One can see the ITG mode in the lower $k_y\rho_i$ range of Fig.
\ref{fig:BenchmarkITG} characterized by propagation in the $\omega >0$ ion diamagnetic drift direction. The ETG mode becomes unstable at higher $k_y\rho_i$ where $\omega$
crosses to the negative electron diamagnetic drift direction \cite{HortonTransport}. While the CTEM is expected to contribute to electron transport \cite{BatchTrans,JenkoTEM,
XiaoTEM,ChenCTEM2}, Fig. \ref{fig:GrowthRatesTEM} shows that it is stable in the intermediate-scale range.
   \begin{figure*}[!ht]
      \label{fig:BaseETG}
      \centering
      \subfigure[]{
      \label{fig:NonlinearHeatFlux}
      \includegraphics[width=2.2in]{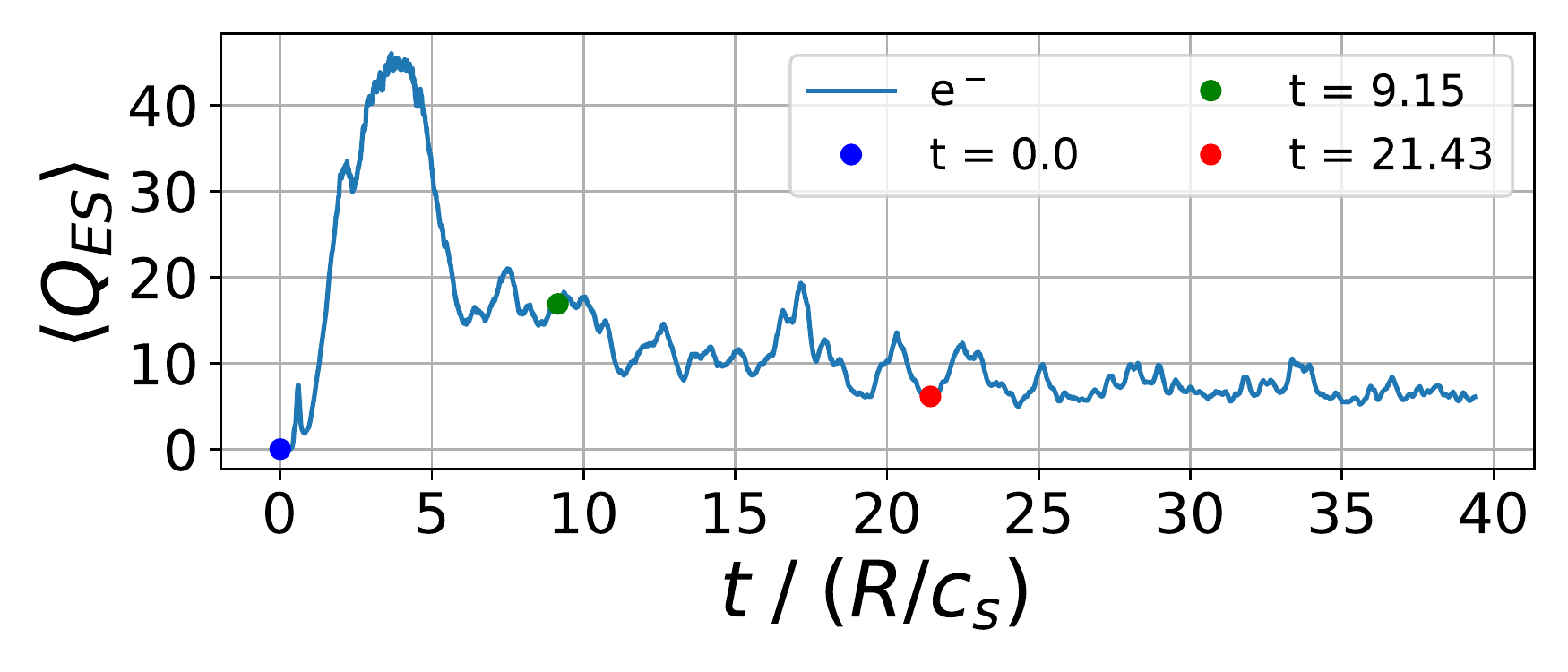}}
      \subfigure[]{
      \label{fig:EarlyRealPhi}
      \includegraphics[width=2.3in]{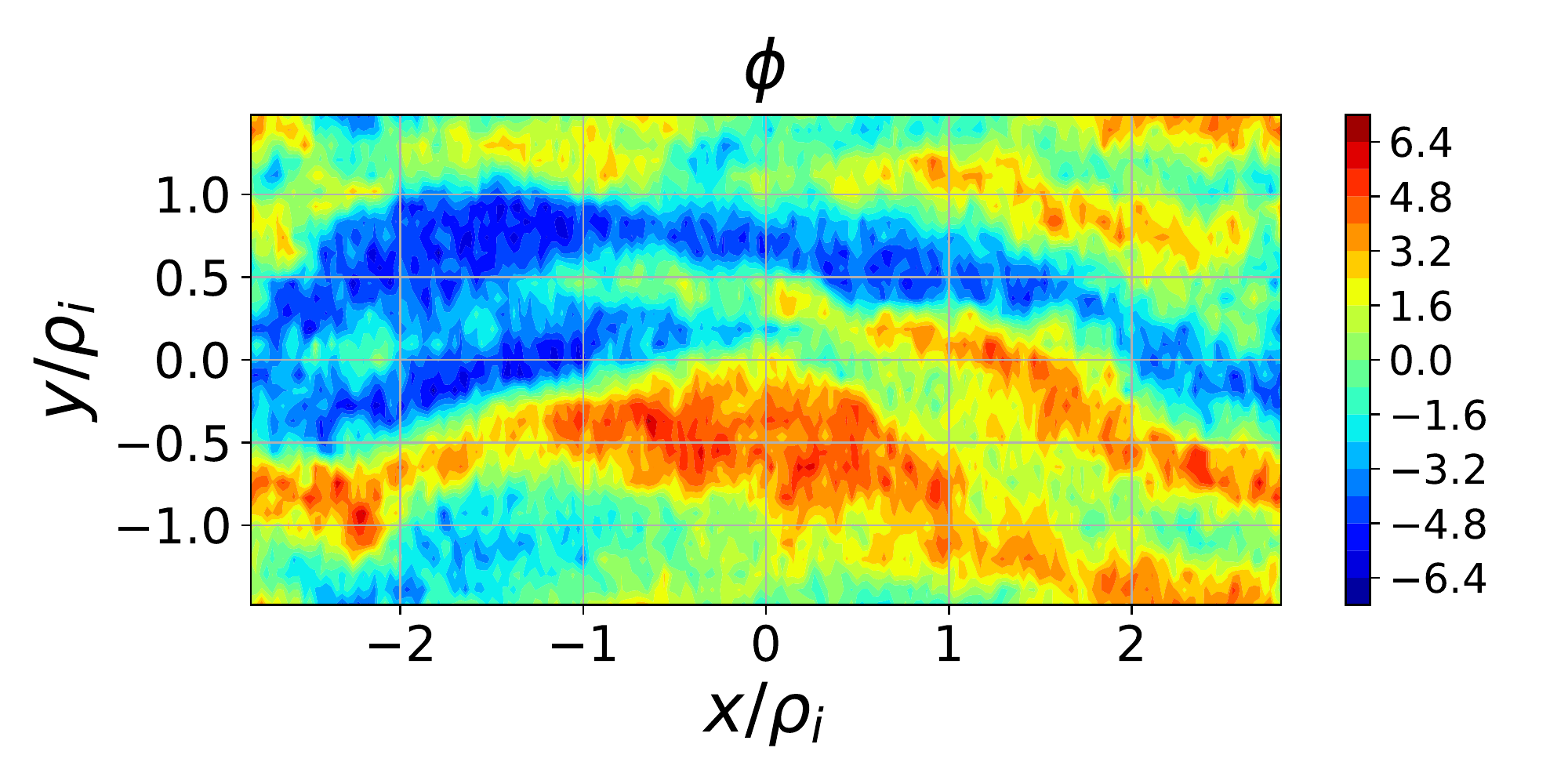}}
      \subfigure[]{
      \label{fig:LateRealPhi}
      \includegraphics[width=2.3in]{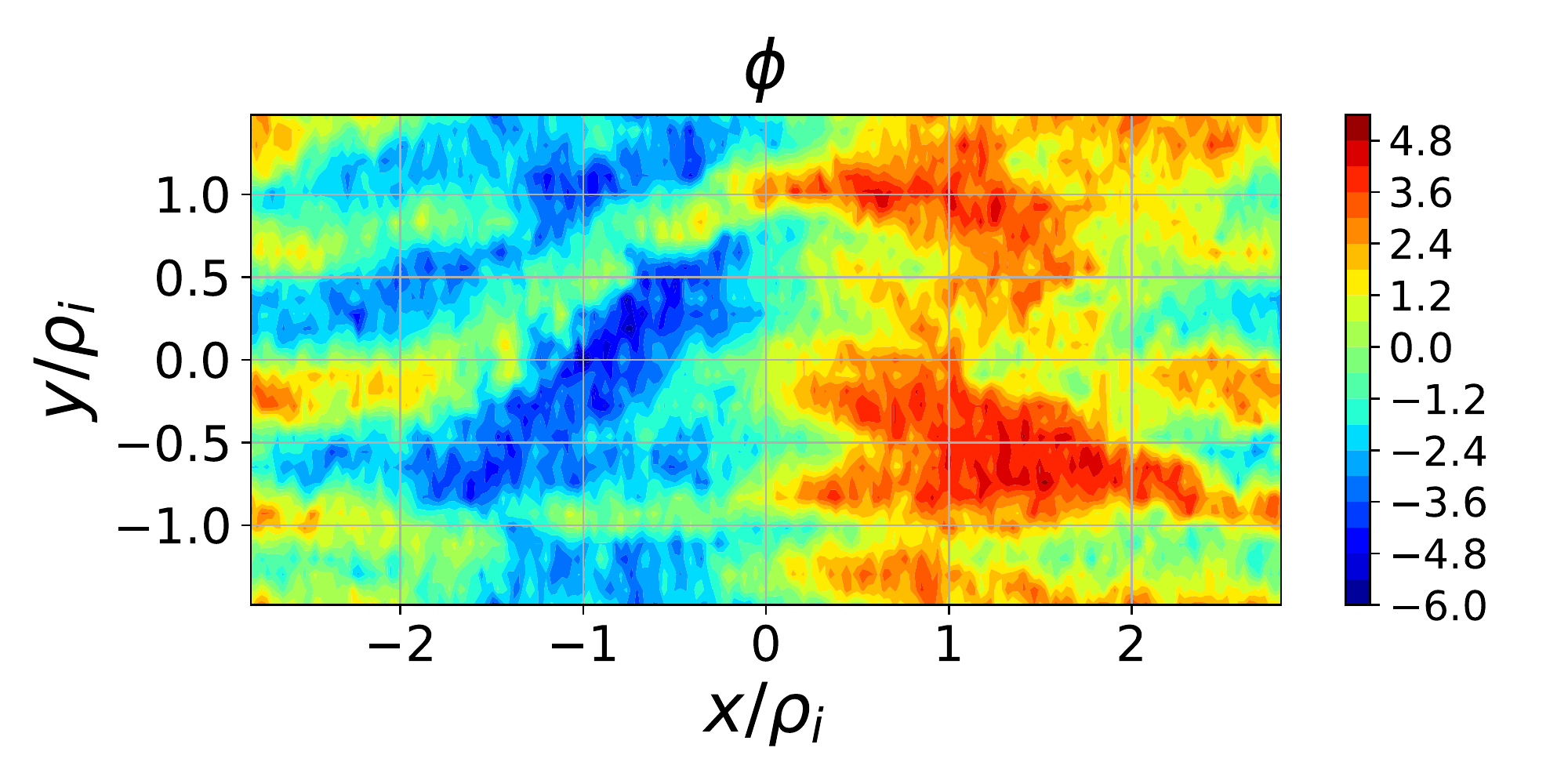}}
      \caption{Original (small-box, no collisions) nonlinear simulation results showing (a) time-marked electrostatic electron heat flux and electrostatic potential
               contours for the (b) early nonlinear phase (green marker), and (c) late zonal phase (red marker).}
   \end{figure*}
   \begin{figure*}[!ht]
      \centering
      \subfigure[]{
      \label{fig:InitFourierPhi}
      \includegraphics[width=2.2in]{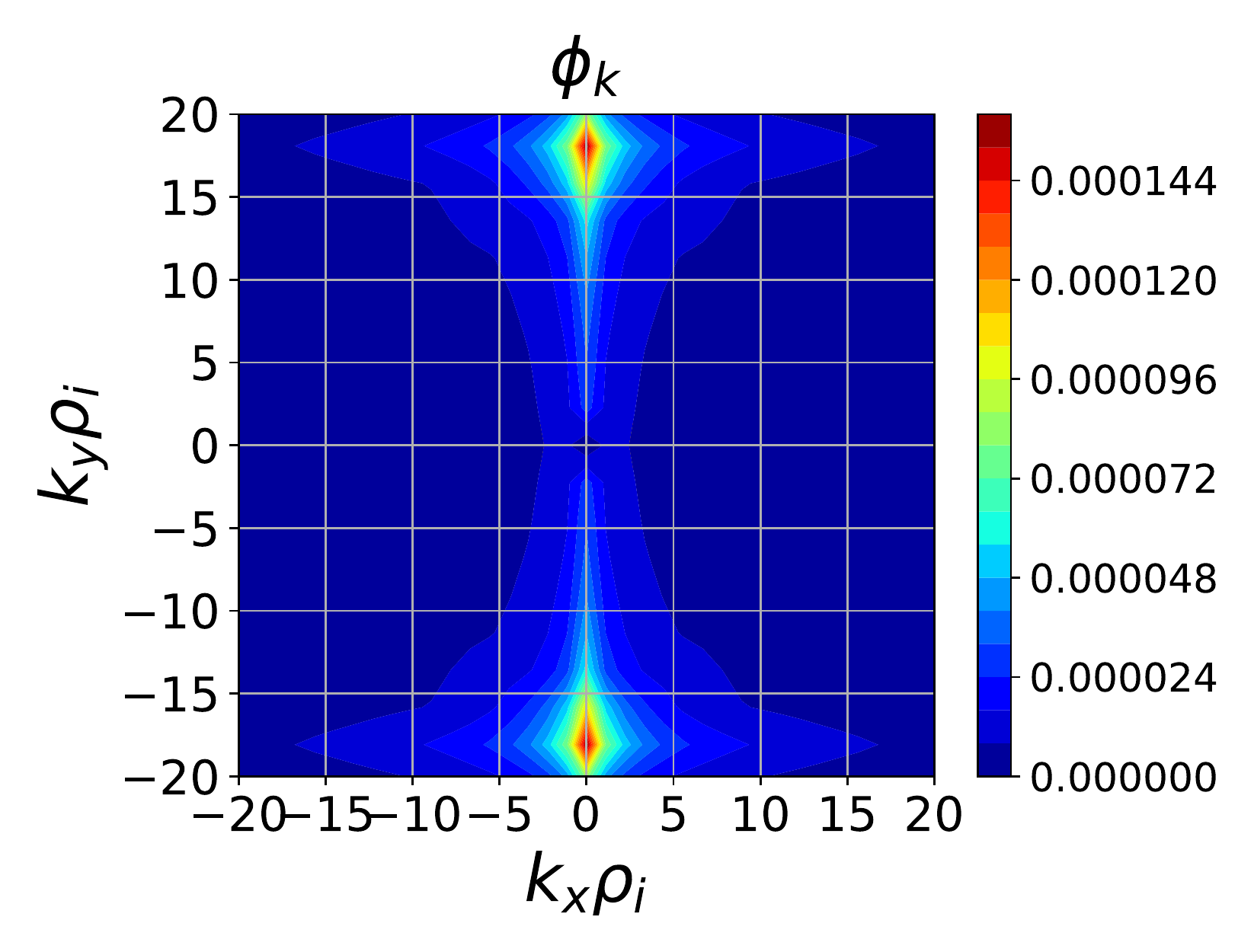}}
      \subfigure[]{
      \label{fig:EarlyFourierPhi}
      \includegraphics[width=2.2in]{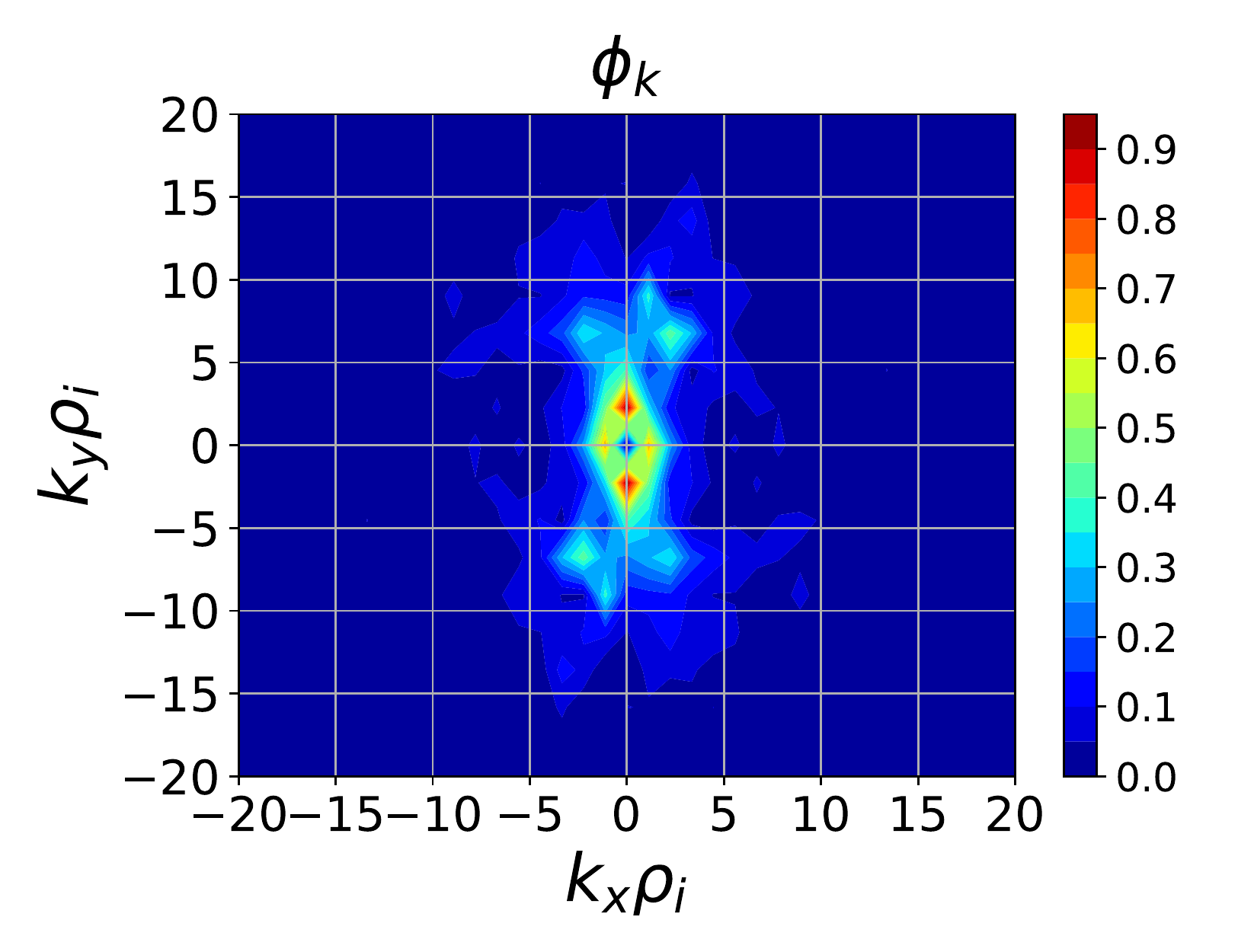}}
      \subfigure[]{
      \label{fig:LateFourierPhi}
      \includegraphics[width=2.2in]{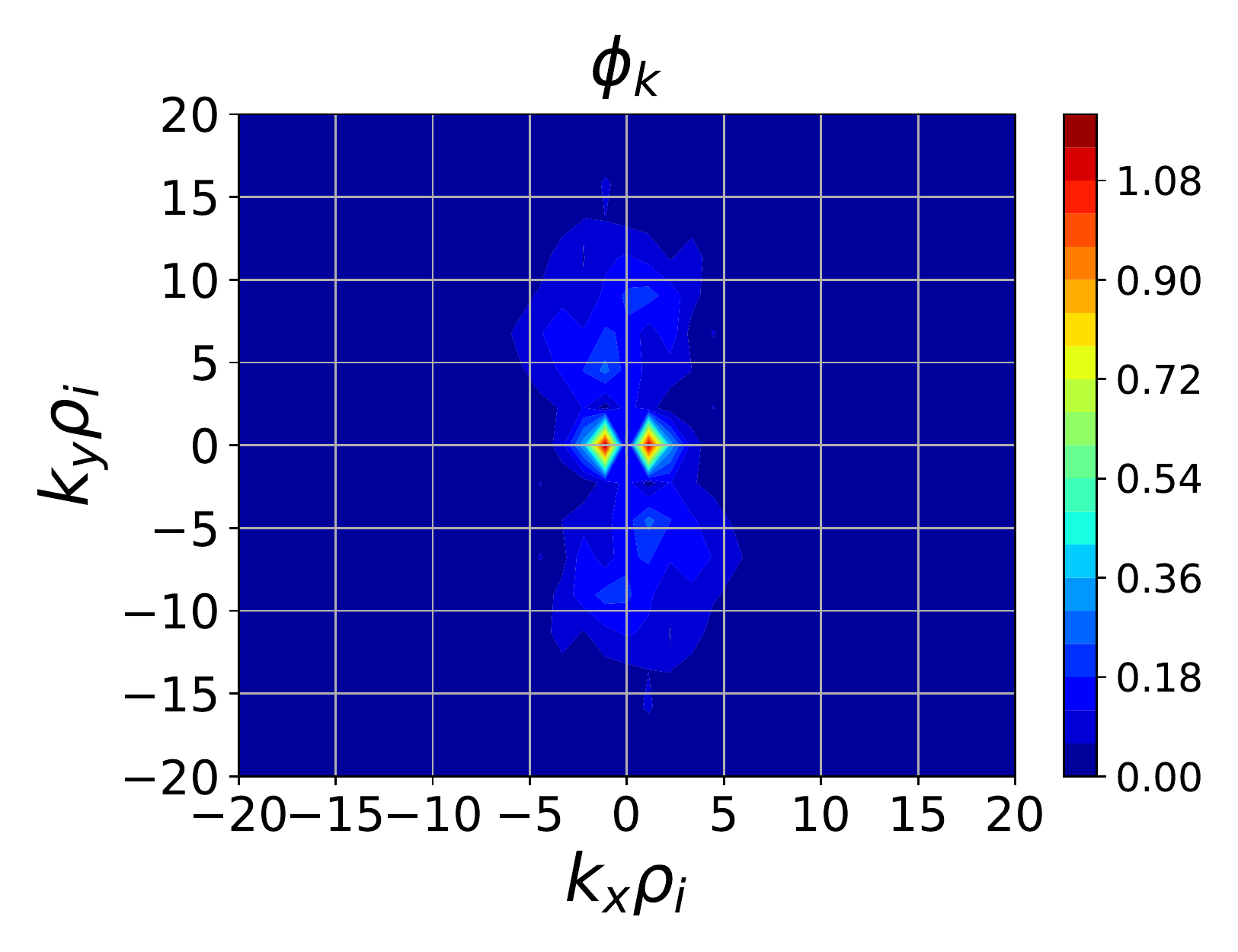}}
      \caption{Original (small-box, no collisions) nonlinear simulation results showing Fourier-space electrostatic potential contours for the (a) initial time
              (blue marker), (b) early nonlinear phase (green marker), and (c) late zonal phase (red marker). The markers correspond to the times marked in Fig.
              \ref{fig:NonlinearHeatFlux}.}
   \end{figure*}

The reference values and radial profiles are taken as specified in Ref. \cite{GorlerBenchmark} with reduced values of $m_i$ and $\beta=8\pi n_eT_e/B^2$. Here $n_e$ and
$T_e$ are the electron density and temperature, and $B$ is the on-axis magnetic field. $\beta=10^{-5}$ was chosen to keep the simulation nearly electrostatic in order to
avoid transport due to electromagnetic fluctuations. The safety factor, $q=rB_{\phi}/RB_{\theta}$, and magnetic shear, $\hat{s}$, profiles are given by  $q(r) = 2.52(r/a)^2
- 0.16(r/a) + 0.86$ and $\hat{s} = \frac{r}{q}\frac{dq}{dr}$ \cite{GorlerBenchmark}. $B_{\phi}$ and $B_{\theta}$ represent the toroidal and poloidal magnetic field components
respectively. The radial flux-surface coordinate $r=0.5a$ was chosen, where $a$ is the tokamak minor radius. The normalized density and temperature gradient profiles, for
a general profile $A(r)$, are defined as $R/L_A = -R\partial_r\textrm{ln}(A(r))$, which can be calculated using the profile given in Eq. (2) of Ref. \cite{GorlerBenchmark}.
The values of the pertinent simulation parameters are listed in Table \hyperref[tableTwo]{II}.
   \begin{table}[!b]
      \centering
      Table I. Relevant GENE normalizations and definitions \cite{GeneCode}.
      \begin{center}
      \begin{tabular}{p{.5\linewidth} c}
         \hline
         \hline
         $\rho^*$             & $\rho_i/R$            \\
         $c_s$                & $\sqrt{T_i/m_e}$      \\
         $t$                  & $R/c_s$               \\
         $\omega$             & $c_s/R$               \\
         $\gamma$             & $c_s/R$               \\
         $Q_{\textrm{gB}}$    & $c_sn_eT_e(\rho^*)^2$ \\
         $\chi_{\textrm{gB}}$ & $\rho_i^2c_s/R$       \\
         $\phi$               & $e\phi/(T_e\rho^*)$   \\
         $\nu_c$              & $\pi\textrm{ln}\Lambda e^4n_eR/(2^{3/2}T_e^2)$ \\
         $\Lambda$            & $24 - \textrm{ln}(\sqrt{10^{13}\cdot n_e}/(10^3T_e))$ \\
         \hline
         \hline
      \end{tabular}
      \end{center}
      \label{tableOne}
   \end{table}
   \begin{table}[!b]
      \centering
      Table II. GENE simulation parameters.
      \vspace*{-.5cm}
      \begin{center}
      \begin{tabular}{p{.8\linewidth} c}
         \hline
         \hline
         $R(\textrm{m})$                   & $1.67$                 \\
         $n_{i,e}(10^{19}\textrm{m}^{-3})$ & $4.66$                 \\
         $T_{i,e}(\textrm{keV})$           & $2.14$                 \\
         $B_{\phi}(\textrm{T})$            & $2.0$                  \\
         $r/a$                             & $0.5$                  \\
         $a/R$                             & $0.36$                 \\
         $\rho^*$                          & $0.001413$             \\
         $\beta$                           & $1\textrm{e}{-4}$      \\
         $m_i/m_p$                         & $1.0$                  \\
         $m_e/m_i$                         & $5.4462\textrm{e}{-4}$ \\
         $R/L_{T_i}$                       & $0$                    \\
         $R/L_{T_e}$                       & $6.96$                 \\
         $R/L_{n_{i,e}}$                   & $2.22$                 \\
         $q$                               & $1.41$                 \\
         $\hat{s}$                         & $0.837$                \\
         $\nu_{ei}$                        & $0.106875$             \\
         \hline
         \hline
      \end{tabular}
      \end{center}
      \label{tableTwo}
   \end{table}

\quad For the linear ITG case, the grid resolution was taken with $32$ grid points in the radial dimension, $x$, and $16$ grid points in the parallel spatial dimension, $z$.
The GENE radial coordinate $x$ corresponds to the flux-surface coordinate $r$ for the case of a circular geometry. In Fig. \ref{fig:BenchmarkITG} the flux-tube GENE benchmark
result from Ref. \cite{GorlerBenchmark} is marked as ``G\"orler" and the corresponding ``low" ($32\times 8$) and ``high" ($64\times 16$) velocity grid ($v_{\parallel}\times\mu$)
resolution simulations have been plotted collectively. There is good agreement with the benchmark case in the ITG range. The intermediate scale is well resolved in the ``low
v-res" case, and the same grid resolution was used for the nonlinear simulation, but with the radial grid resolution increased to $192$ gridpoints. The perpendicular box size
was reduced from $125\rho_i\times125\rho_i$ at the ITG turbulence scale to $6\rho_i\times 3\rho_i$ at the ETG turbulence scale, where the original radial extent of the flux-tube
domain was increased from $3\rho_i$ to $6\rho_i$ to allow for the full formation of the ETG mode streamers.

\quad The electrostatic portion of the radial heat flux, $\langle Q_{ES} \rangle$, for electrons is shown approaching a statistically steady state in time in Fig.
\ref{fig:NonlinearHeatFlux} for the collisionless, $6\rho_i\times 3\rho_i$, nonlinear, full-spectrum, electron-scale case. The heat flux is normalized to $Q_{gB}$, the gyroBohm
normalization given in Table \hyperref[tableOne]{I}, and the angled brackets, $\langle ...\rangle$, denote a flux-tube volume average. The heat flux is determined in GENE as
\cite{GeneCode},

   \vspace*{-.5cm}
   \begin{equation}
      Q_{ES} = \int d^3v \left (\frac{1}{2} mv^2 \right ) \bm{v}_{E\times B} \cdot \hat{\bf r} \; \delta f \;,
   \end{equation}
where $\delta f$ is the distribution function perturbation, $\bm{v}_{E\times B} = -(\nabla\phi\times\bm{B})/B^2$ is the $\bm{E}\times\bm{B}$ drift, $\phi$ is the electrostatic
potential perturbation, and $m$ and $v$ are the particle mass and velocity. The normalization for $\phi$ is given in Table \hyperref[tableOne]{I}. One can see the shift from the
early, nonlinear state characterized by radially-elongated electrostatic potential streamers in Fig. \ref{fig:EarlyRealPhi} to the later state of Fig. \ref{fig:LateRealPhi} where
zonal flows have become dominant. It is during this phase that intermediate-scale zonal flows grow slowly into the final quasi-saturated state.

\quad The initial condition for $\phi$ was realistically peaked about the most unstable mode as shown in Fig. \ref{fig:InitFourierPhi}. This allowed for a transition from the
high-$k_y$ ETG turbulence regime to the intermediate scale where ZF generation is expected to be stronger \cite{Chen_ETG_ZF}. Such zonal flow generation is not present in toroidal
electron fluid theories \cite{LinFluidETG,ChenFluidETG}. An inverse-cascade can clearly be seen between Figs. \ref{fig:InitFourierPhi} and \ref{fig:EarlyFourierPhi}. This
initial saturation is discussed further for the single-mode simulation results presented in Sec. \ref{sec:SingleModeResults}, and for the well-converged, collisional, full-spectrum
simulation results presented in Sec. \hyperref[sec:FullSpec]{IV}. The convergence tests for finding an optimal nonlinear box size are detailed in Appendix \hyperref[sec:ConvergenceNL]{A}.

\section{\label{sec:SingleMode}Single-Mode Analysis}

\subsection{\label{sec:NLSE_Theory}Zonal Flow Generation Mechanism}

\quad It was shown theoretically in Ref. \cite{Chen_ETG_ZF} that intermediate-scale zonal flow may play a role in the nonlinear saturation of ETG turbulence in tokamak plasmas.
The equations of Ref. \cite{Chen_ETG_ZF} are briefly reviewed here, and the nonlinear single-mode simulation results are then presented in Sec. \ref{sec:SingleModeResults} with
comparison to the theory. The theoretical model takes the standard gyrokinetic equation \cite{FriemanChenGyro} for electrons and the quasineutrality condition in the intermediate-scale
limit $k_{\perp}^2\rho_e^2 \ll 1 \ll k_{\perp}^2\rho_i^2$. For long-wavelength ETG modes, one can generally assume that the growth rate is much smaller than the real frequency,
$\vert\gamma/\omega\vert \ll 1$, which allows for a weak-turbulence analysis. These modes also satisfy the relation $\vert\omega_d\vert \ll \vert\omega\vert,\vert\omega_t\vert$, for
$\omega_t$ and $\omega_d$ the transit and magnetic drift frequencies respectively. Local approximations \cite{KimHortLocal} are then assumed for $\omega_d$ and $\omega_t$ in formulating
a kinetic electron model.

\quad The evolution of a single ETG mode can then be derived in terms of a nonlinear Schrodinger equation (NLSE) \cite{Chen_ETG_ZF},

   \vspace*{-.5cm}
   \begin{equation}\label{NLSE_ETG}
   \begin{aligned}
   \relax[i(\partial_t - \gamma_n) - b_nk_\theta^2\rho_e^2\hat{s}^2\theta_k^2 - \frac{c_n}{k_\theta^2\rho_e^2\hat{s}^2}\frac{\partial^2}{\partial\theta_k^2}]A_n(\theta_k)\\
   = \frac{-ik_\theta\rho_e \hat{s}}{\sqrt{2\pi}} \int d\vartheta_k\vartheta_k A_z(\vartheta_k)A_n(\theta_k-\vartheta_k),
   \end{aligned}
   \end{equation}
with $A_z$ and $A_n$ the amplitudes of the ZF and ETG modes, $\gamma_n$ the ETG mode growth rate, $\theta_k$ and $\vartheta_k$ the tilting angle (defined by $\theta_k = k_x/(k_y\hat{s})$)
in the flux-tube limit \cite{BeerThesis} for $k_y$ of the ETG mode), $k_{\theta}$ the poloidal wavenumber, and the terms including $b_n$ and $c_n$ coming from the frequency mismatch
and plasma nonuniformity corrections respectively. $b_n$ and $c_n$ are specifically associated with the linear ETG dynamics as explained in Ref. \cite{Chen_ETG_ZF}.
The nonlinear term under the integral in Eq. (\ref{NLSE_ETG}) describes a Navier-Stokes type coupling due to \textbf{E}$\times$\textbf{B} shearing effects. This coupling is
$\mathcal{O}((k_{\theta}\rho_e)^{-2})$ stronger than the usual Hasegawa-Mima type coupling in the fluid limit \cite{LinFluidETG,ChenFluidETG}. This stronger coupling results
in a stronger regulation of ETG turbulence by zonal flows and also leads to a reduced threshold for ZF excitation by intermediate-scale ETG modes. The threshold condition
is described further in Eq. (\ref{Threshold}).

\quad The description of zonal flow dynamics is given by the equation \cite{Chen_ETG_ZF},

   \vspace*{-.5cm}
   \begin{equation}\label{NLSE_ZF}
   \begin{aligned}
      \relax[\partial_t + \gamma_z(1+d_zk_\theta^2\rho_e^2 s^2\theta_k^2)]\chi_zA_z(\theta_k)\\
      = \sqrt{\frac{\pi}{2}}(k_\theta\rho_es) \theta_k\int d\vartheta_k\vartheta_k^2[A_n(\vartheta_k)A_n^*(\vartheta_k-\theta_k)a_n^*\\
                                                                                     - A_n(\vartheta_k + \theta_k)A_n^*(\vartheta_k)a_n].
   \end{aligned}
   \end{equation}
The nonlinear term under the integral in Eq. (\ref{NLSE_ZF}) is related to the Reynolds stress of the ETG modes. The ZF damping rate is given by $\gamma_z \simeq 3\nu_{ee}/
(\vert\omega_{*e}\vert\sqrt{\epsilon})$, with $\nu_{ee}$ the electron-electron collision frequency and $\omega_{*e}$ the electron diamagnetic drift frequency. The total
electric susceptibility is defined as $\chi_z = \tau + (1+1.6q^2/\sqrt{\epsilon})k_\theta^2\rho_e^2s^2\theta_k^2/2$ with $\tau = T_e/T_i$. The $d_z$ term in Eq. (\ref{NLSE_ZF})
represents a gyrodiffusive correction which helps suppress short wavelength zonal flows \cite{GyroDiffDampZF}.

\quad The $a_n$ term describes the parallel correlation of the ETG turbulence and is defined as $a_n(\theta_k, \vartheta_k) = \int d\eta\langle\tilde{\Phi}_n^*(\eta,\vartheta_k)
v_{\perp}^2\delta\tilde{H}_n(\eta,\vartheta_k + \theta_k)\rangle_v$, where $\eta$ is the extended poloidal angle. Here, $\delta\tilde{H}_n$, is the non-adiabatic part of
the distribution function perturbation. Keeping $a_n$ non-local in tilting angle in order to take into account ballooning effects leads to a parallel decoupling of ETG
modes and therefore ZFs as well, where $a_n = a_n(0,0)\textrm{exp}(-\theta_k^2/2\bar{\eta}^2)$ for $\bar{\eta} = (\int d\eta\tilde{\Phi}^*\eta^2\tilde{\Phi})^{1/2}$ the
parallel-mode-averaged potential \cite{Chen_ETG_ZF}.

\quad Eqs. (\ref{NLSE_ETG}) and (\ref{NLSE_ZF}) are taken together as the NLSE model. The numerical solution of the NLSE model with a single ETG mode and a range of ZF
modes gives an evolution of ETG and ZF modes that can be described by three specific stages \cite{Chen_ETG_ZF}. The initial stage involves uninhibited exponential growth
of the ETG mode to a threshold point at which the radial beating of the ETG drives ZF growth as described by equation (\ref{NLSE_ZF}). As the ZF modes grow, they lead to
radial dispersion of the initial ETG wave packet and the creation of sidebands. These sidebands then drive more zonal flows via a modulational instability in the second
stage. Once the zonal flow grows to appreciable levels in comparison with the ETG mode, the nonlinear interaction in equation (\ref{NLSE_ETG}) acts to saturate the ETG
mode. In the final stage, the linear growth rate of the ETG mode becomes negligible and the NLSE model then results in slow, algebraic growth for the zonal flow
\cite{Chen_ETG_ZF}. This slow growth has been observed in previous gyrokinetic electron-scale simulations \cite{ParkerZF,ColyerETG}.

\quad A threshold condition for the ZF excitation can be calculated analytically \cite{Chen_ETG_ZF} by considering a simple four-wave model for a single zonal flow mode,
an ETG pump mode, and two ETG sideband modes. Narrow-band, rectangular functions are used to describe the ZF and ETG modes, $A_z\Pi[(\theta_k-\theta_z)/W]$ and $A_0\Pi(\theta_k/W)
+ A_+\Pi[(\theta_k-\theta_z)/W] + A_-\Pi[(\theta_k+\theta_z)/W]$ respectively. Here $W$ is the full-width of the modes, $\theta_z$ is the ZF wavenumber in terms of tilting
angle, and $A_z, A_0,$ and $A_{\pm}$ are the ZF, ETG pump, and ETG sideband mode amplitudes. Substituting these functions into Eqs. (\ref{NLSE_ETG}) and (\ref{NLSE_ZF}) with
the assumption of no plasma nonuniformities ($c_n = 0$) and a steady state pump amplitude for simplicity, one obtains the critical threshold condition \cite{Chen_ETG_ZF},

   \vspace*{-.5cm}
   \begin{equation}\label{Threshold}
      W^2\vert A_{0,c}^2\vert = \frac{(\Delta^2+\gamma_s^2)\gamma_z (1+d_z(k_{\theta}\rho_e\hat{s}\theta_z)^2)\chi_z}{(k_{\theta}\rho_e\hat{s}\theta_z)^4[\gamma_s
      \textrm{Re}(a_n) + \Delta \textrm{Im}(a_n)]}.
   \end{equation}
Here, $\Delta = \textrm{Re}(b_n)(k_{\theta}\hat{s}\theta_z\rho_e)^2$ represents the frequency mismatch of the pump and sidebands, and $\gamma_s = \gamma_n + \textrm{Im}
(b_n)(k_{\theta}\hat{s}\theta_z\rho_e)^2$ is the growth rate of the sidebands. This threshold condition for ZF excitation by intermediate-scale ETG modes is $\mathcal{O}
(k_{\theta}^2\rho_e^2)$ lower than the condition found in the fluid approximation \cite{ChenFluidETG}, which would then lead to more effective ZF generation at intermediate
scales.
   \begin{figure*}[!ht]
      \label{fig:SingleModeComp}
      \centering
      \begin{minipage}{\textwidth}
         \includegraphics[width=.9\linewidth]{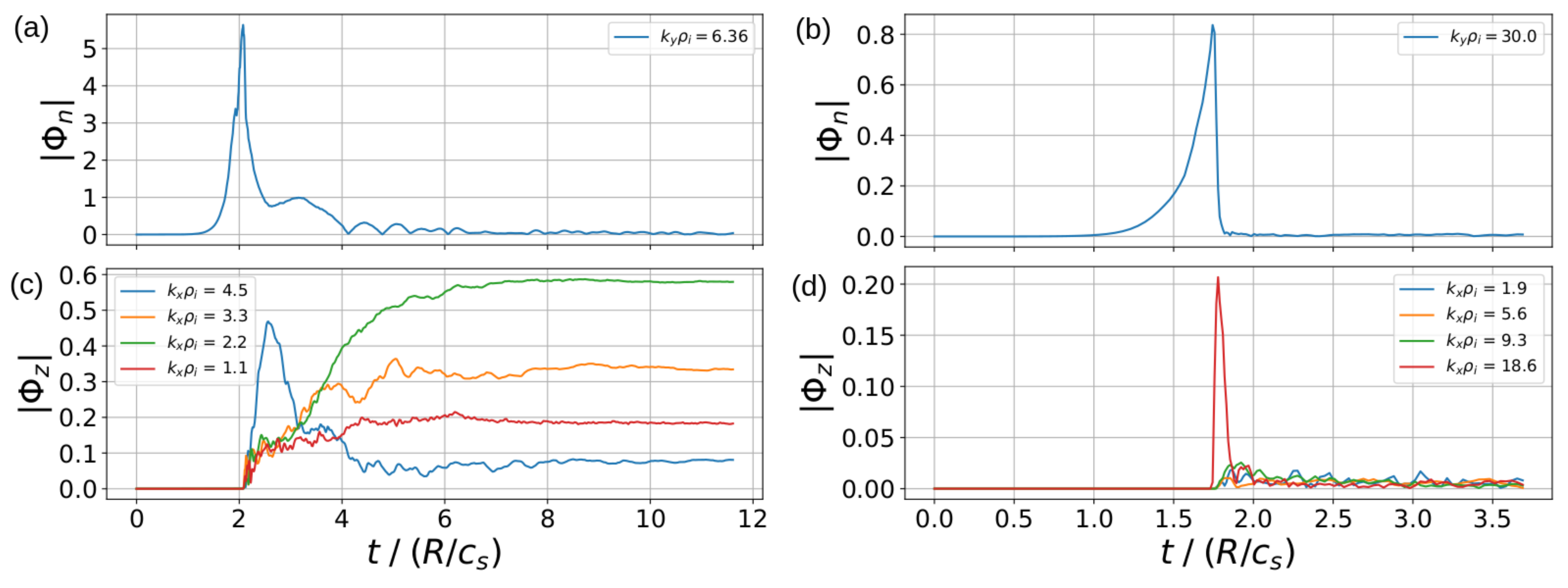}
         \caption{Plots of collisionless single ETG mode evolution for (a) $k_y\rho_i=6.36$ and (b) $k_y\rho_i=30$ with respective growth rates $\gamma\approx 7.037$ and
                  $7.015$. The strongest four ZF modes in the late time are plotted for (c) the $k_y\rho_i=6.36$ case, while (d) shows the excitation of a larger range of
                  ZF modes in the $k_y\rho_i=30$ case.}
      \end{minipage}
   \end{figure*}
   \begin{figure}[!b]
      \centering
      \includegraphics[width=.8\linewidth]{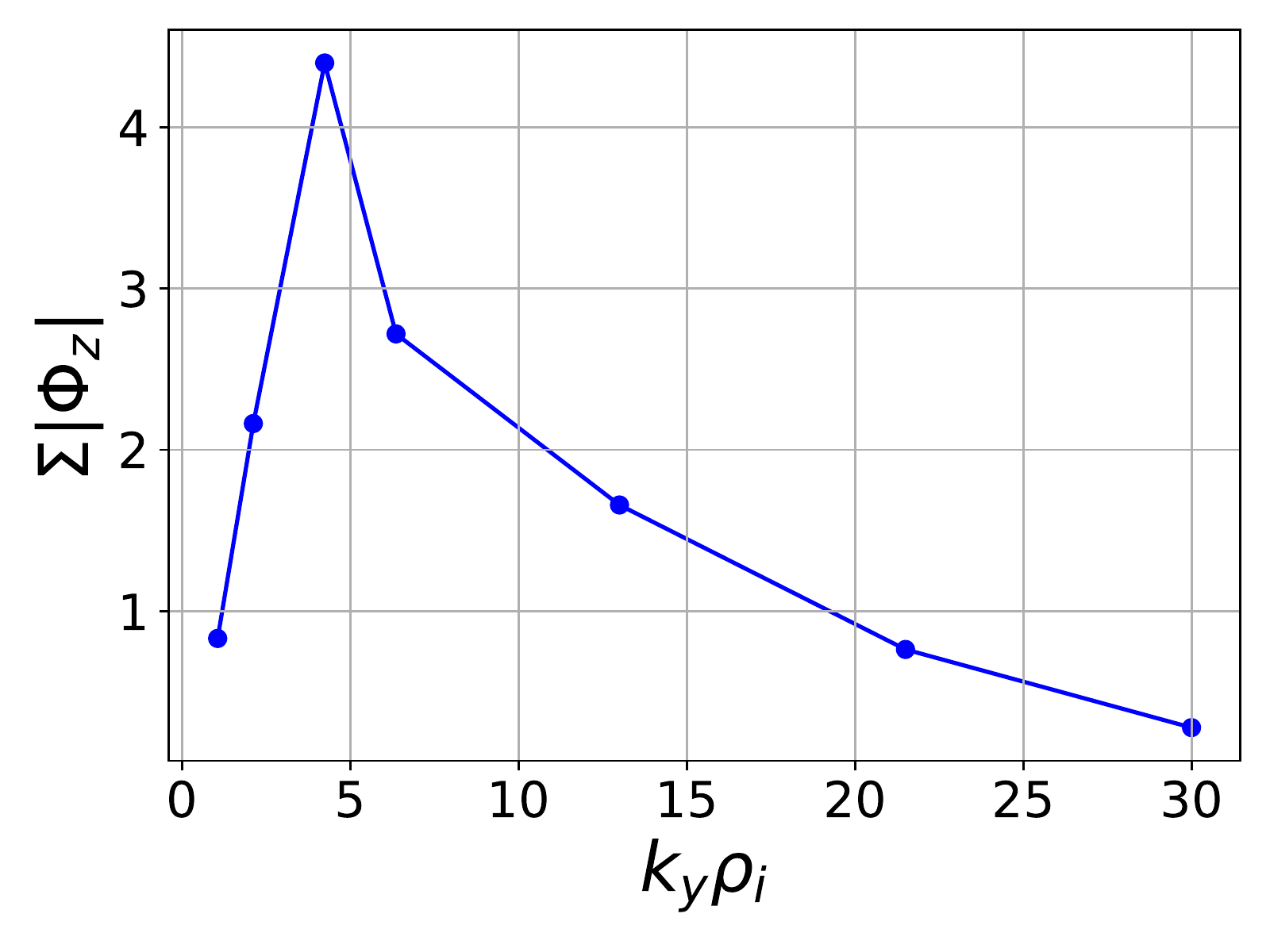}
      \caption{Plot of the sum of zonal potential amplitudes at final times, $\sum\vert\phi_z\vert = \sum_{k_x}\vert\phi_{k_x,0}\vert$, as driven by a single ETG mode.
               The fourth and seventh points correspond to the ETG modes from Fig. \hyperref[fig:SingleModeComp]{4}.}
      \label{fig:PhiTotScan}
   \end{figure}
   \begin{figure*}[!ht]
      \centering
      \begin{minipage}[!t]{.4\textwidth}
         \label{fig:TimeTraceTot}
         \includegraphics[width=\linewidth]{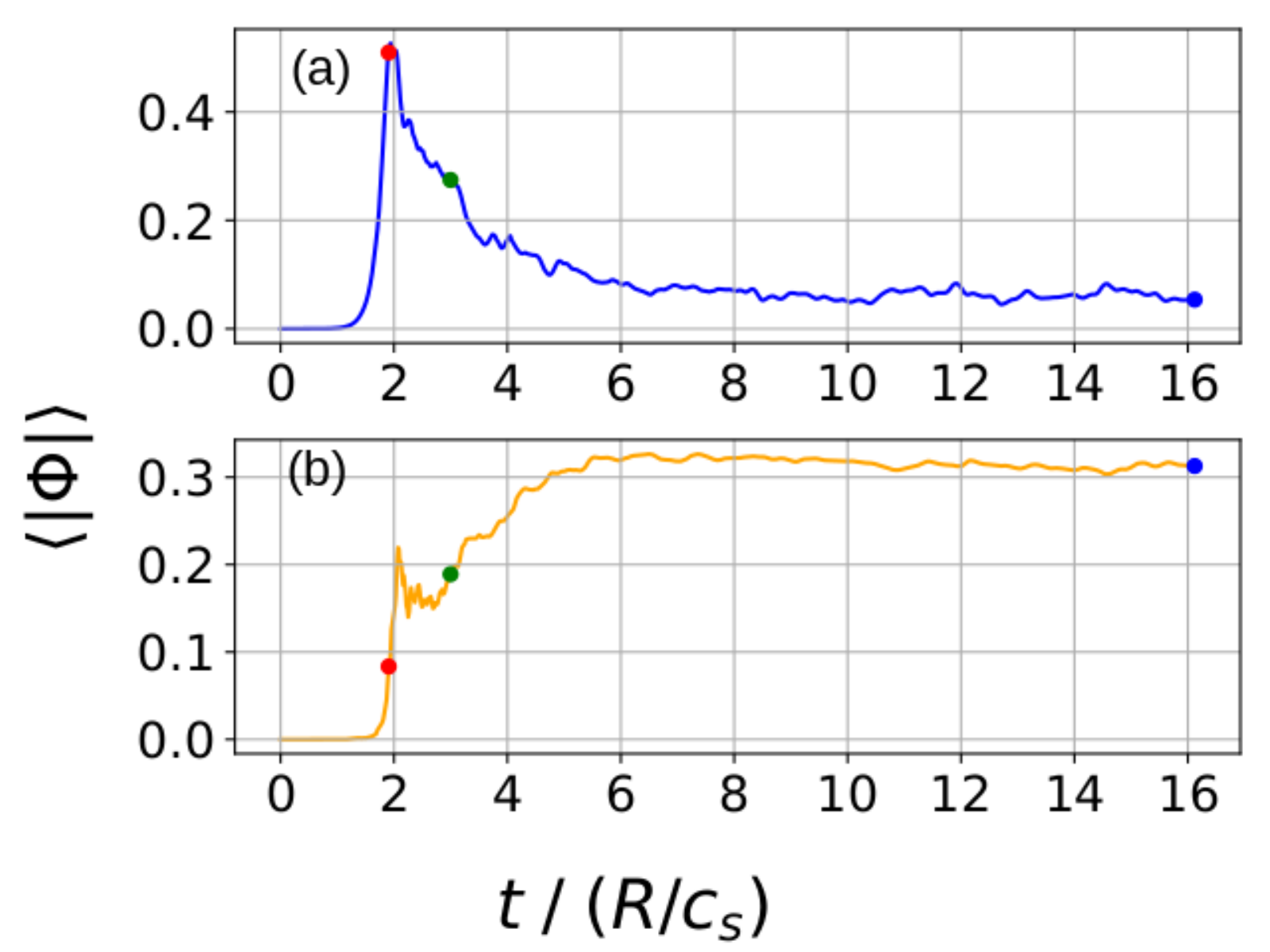}
         \caption{Plot of total (a) ETG and (b) ZF mode potential as an integral over $k_x$ in the collisional, Gaussian-$k_x$ ETG case for $k_y\rho_i=6.36$. Markers
                  have been added to match the spectral snapshots of Fig. \ref{fig:SingleMode6_SpecEv}.}
      \end{minipage}%
      \hspace*{.5cm}
      \begin{minipage}[!t]{.6\textwidth}
         \vspace*{.55cm}
         \centering
         \includegraphics[width=\linewidth]{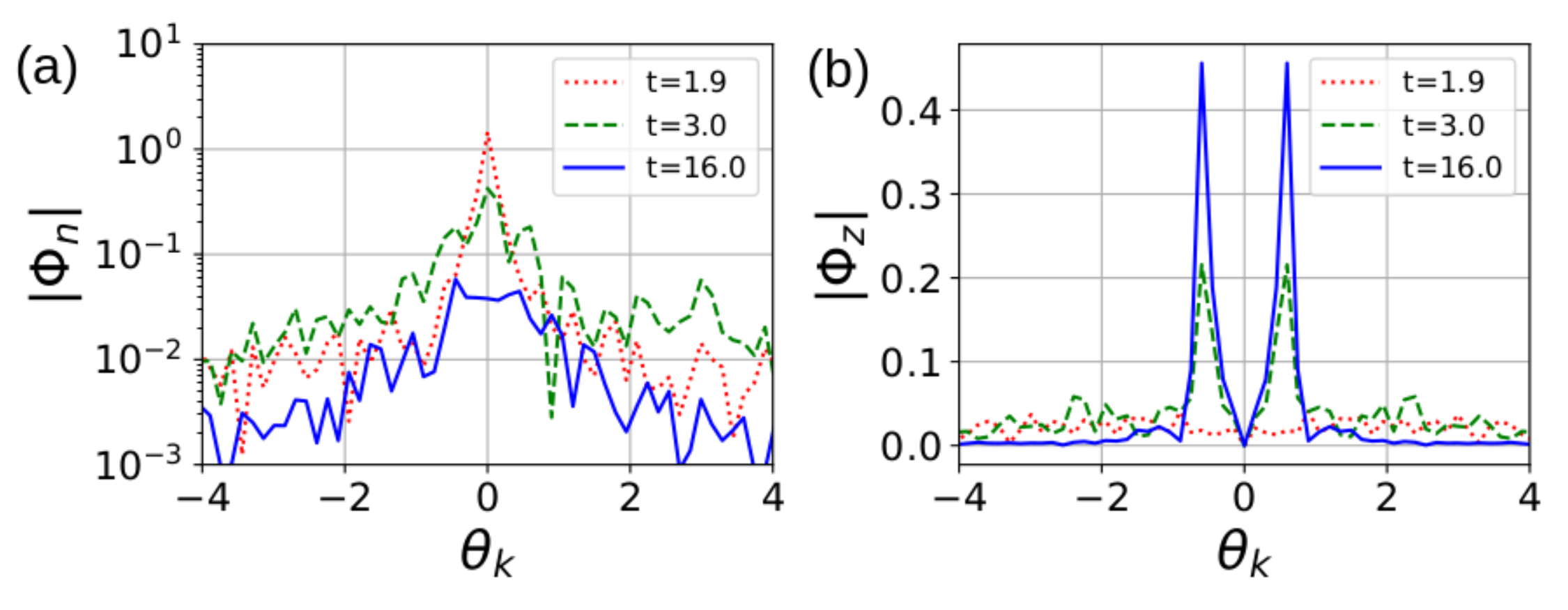}
         \caption{Radial-spectrum snapshots of (a) the single ETG mode and (b) ZF as a function of tilting angle for the $k_y\rho_i = 6.36$ case. Snapshot times match
                  the markers in Fig. \hyperref[fig:TimeTraceTot]{6}.}
         \label{fig:SingleMode6_SpecEv}
      \end{minipage}
   \end{figure*}

\quad The ZF modes are initially excited in a range of $\theta_z$ values. As the system evolves to the quasi-saturated state and the ETG mode is suppressed, the ZF spectrum
narrows towards the most easily driven mode. Comparing the exponential $\theta_k$ dependence of the $a_n$ term to the algebraic form of the $d_z$ term, one finds that the
parallel decoupling is largely responsible for minimizing the intermediate-scale threshold condition at low $\theta_z$. If one considers the temporal evolution of the four-wave
model, a fixed-point solution can be found with constant $A_z$ and sufficiently low ZF damping rate, where the ZF and ETG mode amplitudes are given as \cite{Chen_ETG_ZF},

   \vspace*{-.5cm}
   \begin{equation}\label{SatAmpZF}
      W^2\vert A_{z,p}\vert^2 = \frac{\pi(\delta^2 - \Delta\delta - \gamma_n\gamma_s)}{(k_{\theta}\hat{s}\theta_z\rho_e)^2},
   \end{equation}
and,

   \vspace*{-.5cm}
   \begin{equation}\label{SatAmpETG}
      W^2\vert A_{0,p}\vert^2 = \frac{\chi_z\gamma_z(1+d_zk_{\theta}^2\hat{s}^2\theta_z^2\rho_e^2)[\gamma_s^2 + (\delta - \Delta)^2]}
      {(k_{\theta}\hat{s}\theta_z\rho_e)^4[(\Delta-\delta)\textrm{Im}(a_n)+\gamma_s\textrm{Re}(a_n)]}.
   \end{equation}
Here, $\delta = \Delta\gamma_n/(\gamma_s + \gamma_n)$ represents the amplitude oscillation frequency of ETG modes due to the nonlinear ETG-ZF coupling. The single ETG
mode spectrum then continues to fluctuate in $k_x$ while the ZF mode reaches a constant, steady state \cite{Chen_ETG_ZF}. Additionally, $\vert A_{0,p}\vert^2$ is then
proportional to $\gamma_z$ in the saturated state, and therefore the collision frequency, while $\vert A_{z,p}\vert^2$ is not. While these saturation estimates are only
valid for a single ZF mode, as the ETG turbulence saturates and the ZF spectrum narrows due to the threshold condition, $\theta_z$ of the most optimally-driven mode can
be used to estimate the ETG saturation level.

\subsection{\label{sec:SingleModeResults}Single-Mode Simulation Results}

\quad We now compare the NLSE model (Eqs. (\ref{NLSE_ETG}) and (\ref{NLSE_ZF})) to gyrokinetic simulation results. Collisionless, nonlinear, ETG simulations were carried
out where a single unstable ETG mode ($k_x = 0$) and all zonal flow modes ($k_y=0$) are retained. This fairly accurately describes the dynamics of the NLSE model. All
results presented in this section are averaged over $z$. The ETG growth rate spectrum with respect to $k_y\rho_i$ can be utilized here to illustrate the NLSE model dynamics.
As seen in Fig. \ref{fig:GrowthRatesETG}, the ETG growth rate spectrum is symmetric around the most unstable mode, so one can choose to compare the evolution of a pair of
ETG modes with similar growth rates, where one mode has a $k_y\rho_i$ value in the intermediate-scale range and the other mode has a larger $k_y\rho_i$ value outside of
that range. Then the ZF drive of the two modes can be compared to verify the expectations from the NLSE model.

\quad The $k_y\rho_i=6.36$ mode with a growth rate of $\gamma \approx 7.037$ and the $k_y\rho_i=30$ mode with a similar growth rate of $\gamma\approx 7.015$ are taken
here for comparison. Fig. \hyperref[fig:SingleModeComp]{4(a)} shows the time evolution of the $k_y\rho_i = 6.36$ mode, while Fig. \hyperref[fig:SingleModeComp]{4(b)}
shows the time evolution of the $k_y\rho_i=30$ mode. Fig. \hyperref[fig:SingleModeComp]{4(c)} shows the time evolution of the four strongest zonal flow modes at the
final time step for the intermediate-scale case, whereas Fig. \hyperref[fig:SingleModeComp]{4(d)} shows a large range of ZF modes excited in the high-$k_y$ case in order to
illustrate a difference in the zonal flow response between the two cases. One can see that initially both ETG modes grow exponentially at similar rates until a threshold is
reached, at which point zonal flows are excited. For the intermediate-scale ETG mode, this phase is followed by an algebraically-growing long wavelength ZF phase in which
the ZF modes gradually reach a steady state value. The high-$k_y$ results show no slowly-growing ZF phase at late times. This difference in ZF generation in the late stage is
consistent with the threshold condition given in Eq. (\ref{Threshold}). The intermediate-scale ETG mode continues to slowly drive zonal flows as it is suppressed to lower levels,
whereas the high-$k_y$ ETG mode does not.

\quad The peak level of the ETG mode is much lower for the high-$k_y$ case than for the intermediate-scale case. This result is not expected from the NLSE model as the shearing
of the ETG mode by the wave-wave coupling should be stronger in the intermediate scale than at higher $k_y$. Zonal flows are also generated earlier in the high-$k_y$ case,
indicating a lower threshold initially, which is inconsistent with the NLSE model. One noticeable difference between the single-mode GENE simulations and the NLSE model
is that the NLSE model only includes the ZF shearing suppression mechanism, whereas the single-mode GENE simulations include other saturation mechanisms. Comparing the zonal
response between the two cases, it is found that the initial ZF excitation shown in Fig. \hyperref[fig:SingleModeComp]{4(d)} is much more abrupt, possibly indicative of a
secondary instability \cite{DorlandJenkoETG1,DorlandJenkoETG2,ETGFourSec}. The change in ETG-ZF dynamics in the single-mode results is found to occur near $k_y\rho_i \sim 15$.
This difference in behavior likely indicates the reason for the transition to the intermediate scale mentioned previously in Sec. \ref{sec:Model}, and this is further discussed
in comparison to the full-spectrum simulation results presented in Sec. \ref{sec:FullSpec}.

\quad Fig. \ref{fig:PhiTotScan} shows the sum of all ZF mode amplitudes, $\sum\vert\phi_z\vert$, as a function of $k_y\rho_i$. Each value of $k_y\rho_i$ represents initializing
with a different unstable ETG mode. The sum is taken at the final simulation time, where the ZF mode amplitudes are nearing steady state levels. The notable region of ZF
generation is clearly seen to be in the intermediate-scale range, as expected by the NLSE model. Shorter ETG mode wavelengths correspond to weaker zonal flows at late times,
in agreement with fluid ETG models. In addition, the drop-off at long ETG mode wavelengths is reasonable due to trapped electron effects at this wavenumber range \cite{ChenCTEM3}.
A validation of this expectation for the full-spectrum simulations is provided in Appendix \hyperref[sec:ConvergenceNL]{A}.

\quad The unstable ETG mode is shown to be suppressed at late times in Figs. \hyperref[fig:SingleModeComp]{4(a)} and \hyperref[fig:SingleModeComp]{4(b)}. The total
amplitude, $\langle\vert\phi\vert\rangle = (\int d\theta_k\vert\phi_k\vert^2)^{1/2}$, of the ETG mode is small in comparison to the zonal flow amplitude. In contrast,
the NLSE model simulation results given in Fig. 3 of Ref. \cite{Chen_ETG_ZF} show that the total ETG and ZF mode amplitudes, $\langle\vert\phi\vert\rangle$, are of
similar strength and fluctuating in the late stage. One reason to expect the strong ETG suppression in the gyrokinetic simulations is the lack of collisionality which would
damp the zonal flow due to the collisional dependence of the $\gamma_z$ term in Eq. (\ref{NLSE_ZF}). Additionally, the NLSE model assumes a Gaussian radial spectrum for
the ETG mode, while the single-mode flux-tube GENE simulations take $k_x=0$ initially. The globally-Gaussian radial distribution of the ballooning modes would lead to more
radial ETG mode overlap, which would then drive more ZF generation as predicted by Eq. (\ref{NLSE_ZF}).
   \begin{figure}[!b]
      \centering
      \label{fig:TimeTraceThetak}
      \includegraphics[width=.8\linewidth]{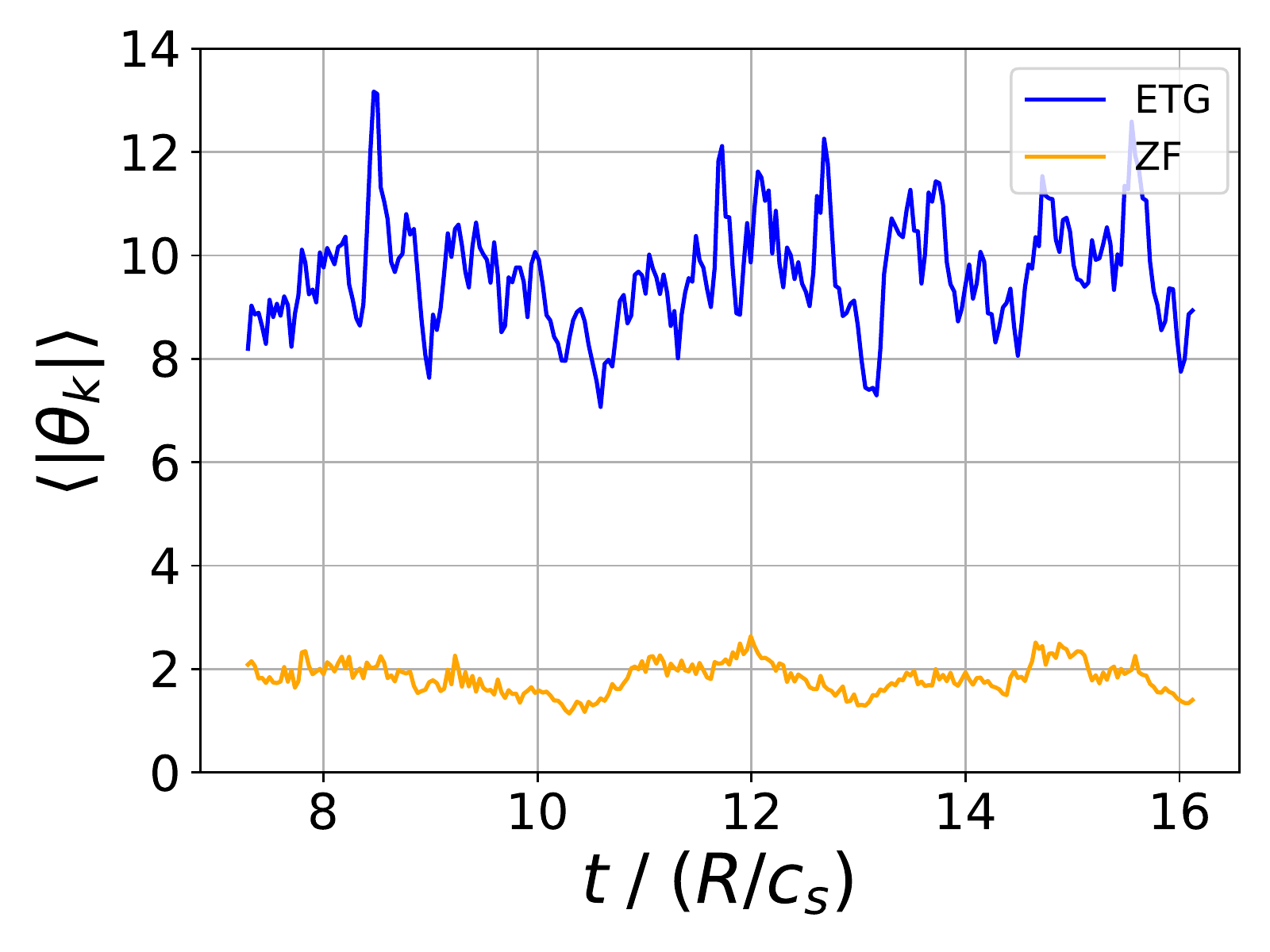}
      \caption{Plot in time of the average dimensionless radial wavenumber for the total ETG and ZF spectra in the quasi-saturated stage.}
   \end{figure}

\quad We were able to obtain more physical results for the $k_y\rho_i = 6.36$ case in which the ETG and ZF modes fluctuate about similar steady state values due to their
nonlinear coupling. These results were achieved by including the physical collisionality taken in Sec. \ref{sec:FullSpec}, such that $\gamma_z = 0.014$, and by initializing
the ETG mode with a Gaussian $k_x$ and $z$ spectrum such that $\phi(k_x,z,t=0) \sim e^{-(k_x^2 + z^2)/8\pi^2}$. The corresponding $\langle\vert\phi\vert\rangle$ for the ETG
and ZF modes are plotted in Fig. \hyperref[fig:TimeTraceTot]{6}. One can see the strong drive of the zonal flow, as well as the late-stage fluctuations of both ETG and ZF
modes. Fig. \ref{fig:SingleMode6_SpecEv} shows that over time the Gaussian radial spectrum of the ETG mode is broadening into sidebands, while the radial spectrum of the
zonal flow modes narrows from a broad distribution to a peak at a final, low-$\theta_z$ mode number, as predicted by the NLSE model \cite{Chen_ETG_ZF}. These results suggest
that one should perform global simulations to see results most consistent with the NLSE model.

\begin{figure}[!b]
      \centering
      \label{fig:AbsAmpRatio}
      \includegraphics[width=.8\linewidth]{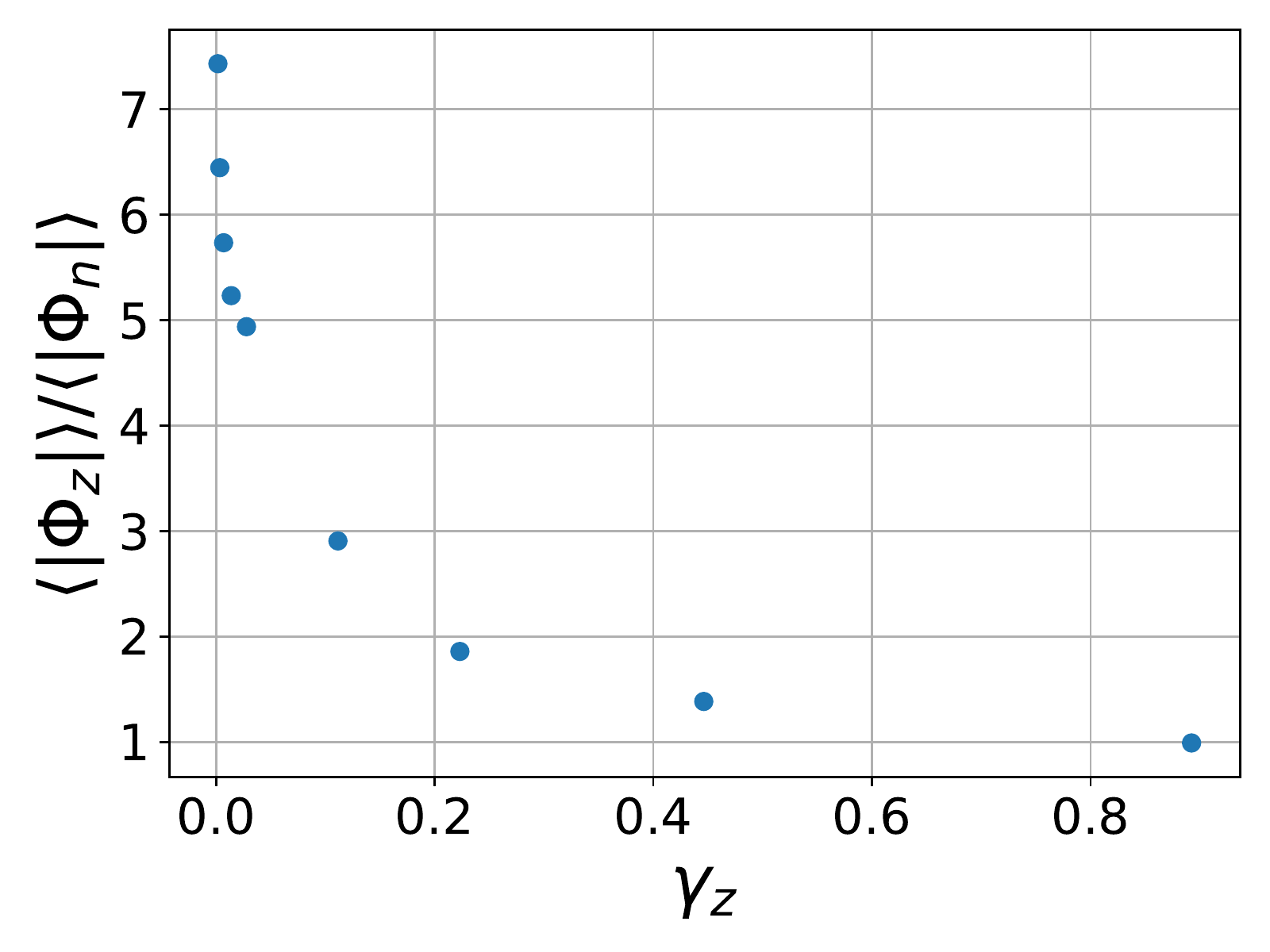}
      \caption{Ratio of total absolute amplitudes of ZF to ETG modes with varying collisionality. Results are taken at the final timestep. The reference value corresponds
      to the fourth point, $\gamma_z = 0.014$.}
   \end{figure}

\quad The improved single-mode results for the $k_y\rho_i=6.36$ case showed large $\theta_k$-averaged fluctuations for the ETG mode compared to more fixed ZF fluctuations. These results agree with
the expectations of Eqs. (\ref{SatAmpZF}) and (\ref{SatAmpETG}). The fluctuations can be seen in Fig. \hyperref[fig:TimeTraceThetak]{8}, where the average dimenionsless radial wavenumber,
$\langle\vert\theta_k\vert\rangle = \left(\int d\theta_k\theta_k^2\vert\phi_k\vert^2 \right)^{1/2}/\langle\vert\phi_k\vert\rangle$, is plotted for the ETG and ZF modes respectively. The
ratio of the total absolute amplitude of ZF to ETG modes is given in Fig. \hyperref[fig:AbsAmpRatio]{9} as a function of $\gamma_z$ and is consistent with the trend from the NLSE model. The
late-stage behavior of the ETG and ZF $k_x$-spectra, as shown in Figs. \ref{fig:SingleMode6_SpecEv} and \hyperref[fig:TimeTraceThetak]{8}, and the collisional behavior of the mode amplitude
ratio agree well with the late-time behavior reported in the electron-scale MAST simulations of Ref. \cite{ColyerETG}.

\section{\label{sec:FullSpec}Full-spectrum Simulation Results}

\begin{figure}[!t]
      \centering
      \label{fig:CollZonal}
      \includegraphics[width=.8\linewidth]{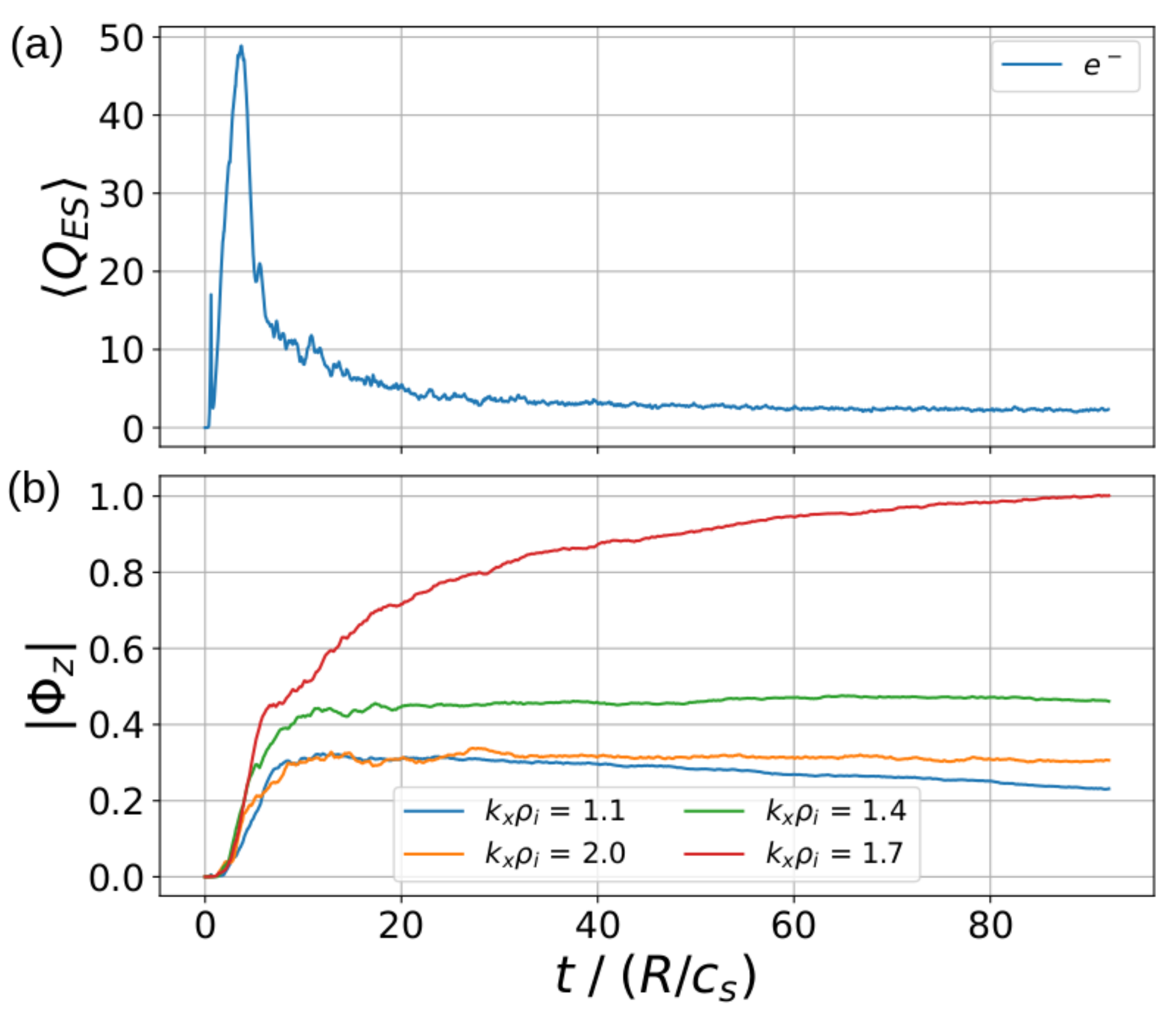}
      \caption{Full-spectrum collisional nonlinear simulation results showing (a) the electron heat flux time-evolution and (b) the time evolution of the four strongest final
               ZF modes. Data is averaged over $z$.}
   \end{figure}

\quad The full-spectrum nonlinear simulation results are presented here and the intermediate-scale zonal flow generation mechanism is further investigated. Including multiple
toroidal modes results in a final quasi-saturated heat flux characterized by richer turbulent interactions. Fig. \hyperref[fig:CollZonal]{10(a)} shows the time history
of the heat flux for the well-converged $24\rho_i\times 3\rho_i$ case with collisionality. The four strongest zonal flow modes at the final time are presented in Fig.
\hyperref[fig:CollZonal]{10(b)}. The value of the normalized electron-ion collision frequency used is $\nu_{ei} = 0.106875$. This frequency is defined as $\nu_{ei} = 4v_{te}\nu_c/R$,
where $v_{te} = \sqrt{T_e/m_e}$ and $\nu_c$ is the collision frequency given in Table \hyperref[tableOne]{I}. The self-adjoint form of the standard Landau-Boltzmann
collision operator is used. Realistic collisionality allows for ZF damping when reaching a final state, and the simulation was carried out to a sufficiently long non-dimensional
time, $t/(R/c_s) = 90$, to ensure that a quasi-saturated steady state in $\langle Q_{ES}\rangle$ is achieved. The convergence with respect to box size is discussed in Appendix
\hyperref[sec:ConvergenceNL]{A}.
   \begin{figure}[!t]
      \label{fig:CollETG}
      \includegraphics[width=.8\linewidth]{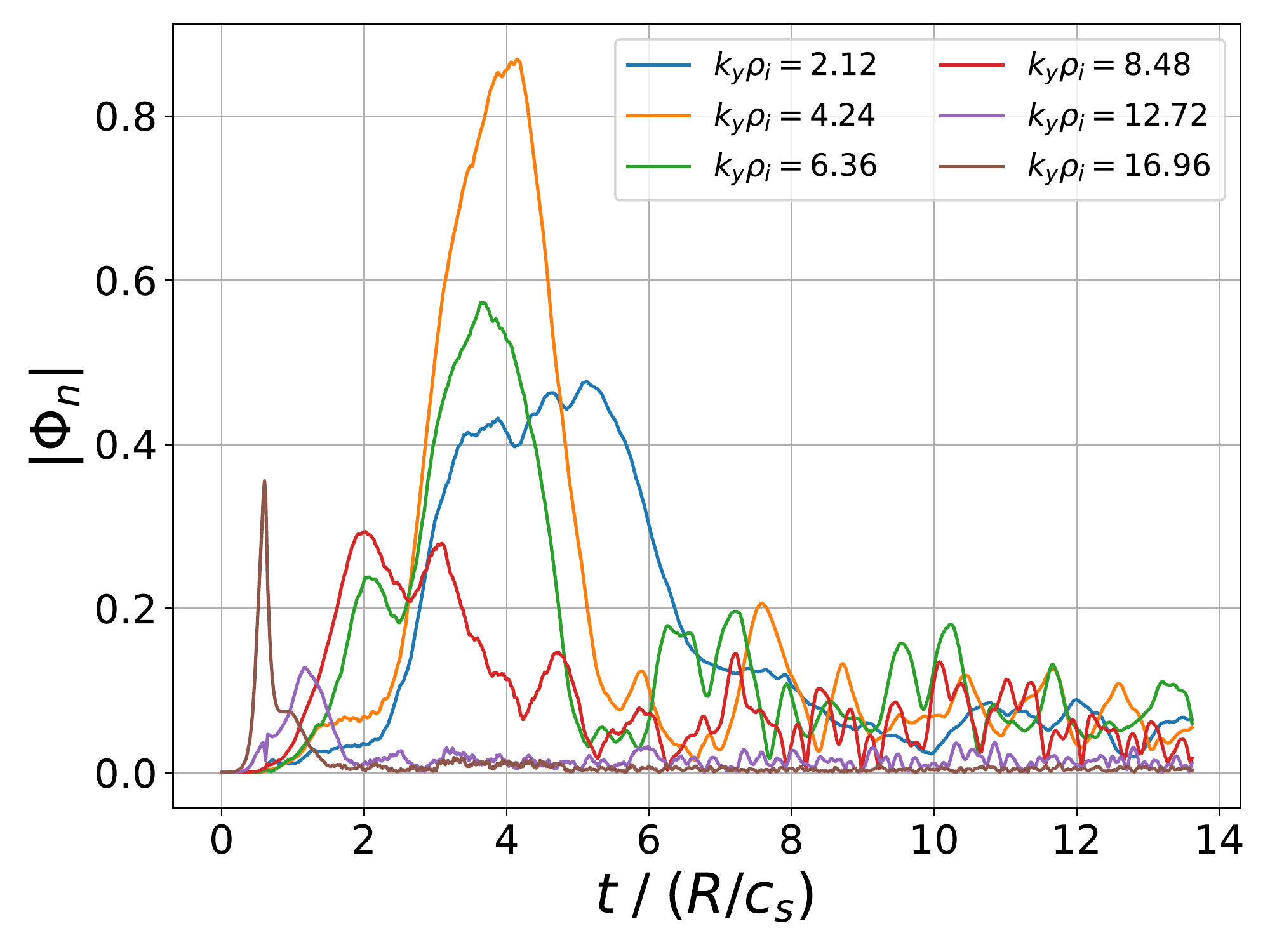}
      \caption{Time trace of full-spectrum ETG modes ranging from the longest mode to the most unstable mode. Data is averaged over $z$.}
   \end{figure}

\quad Fig. \hyperref[fig:CollETG]{11} shows the evolution of various ETG modes ranging from the longest wavelength to the most unstable mode. The shorter wavelength modes
saturate very quickly to negligible levels, in agreement with the single-mode results, and are omitted. One can observe that the intermediate-scale modes saturate the
slowest and reach the highest levels. During the period of intermediate-scale ETG mode growth, the ZF modes shown in Fig. \hyperref[fig:CollZonal]{10(b)} are driven
exponentially by the radial beating of ETG modes, as well as by the modulational instability. Once the ETG modes reach a quasi-saturated state, the ZF modes continue
to grow slowly in agreement with the single-mode simulation results. Considering the findings of the NLSE model, the single-mode simulations results of Sec.
\ref{sec:SingleModeResults}, and the full-spectrum simulation results discussed here, it is the intermediate-scale ETG modes which are most responsible for driving ZF mode
growth into the late stage.
   \begin{figure}[!b]
      \centering
      \label{fig:QuasiSat}
      \includegraphics[width=.9\linewidth, height=5.2cm]{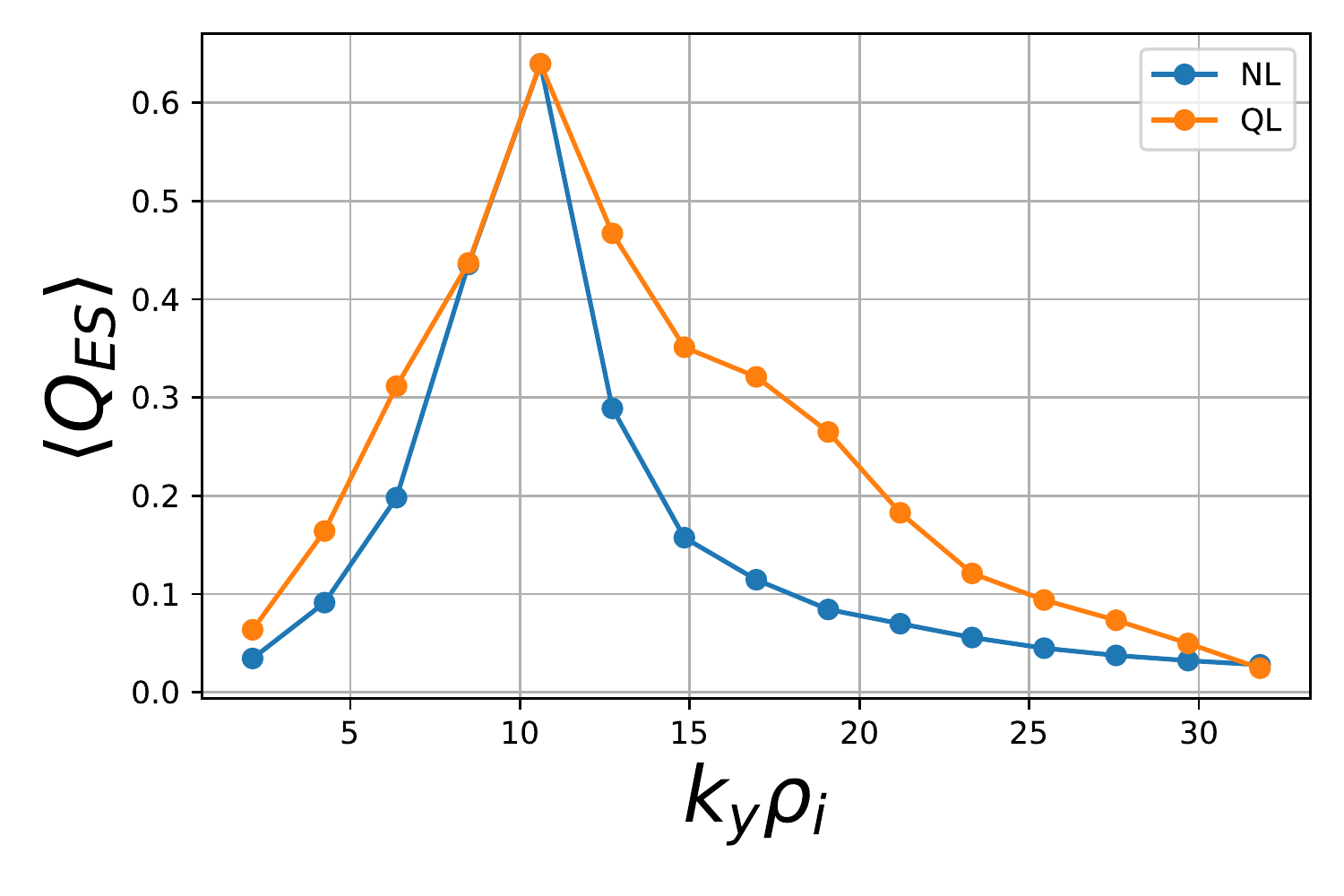}
      \caption{Comparison of quasilinear and nonlinear heat flux spectra for the $24\rho_i\times 3\rho_i$ full-spectrum electron-scale case. NL and QL stand for
               nonlinear and quasilinear respectively.}
   \end{figure}

The heat flux spectrum for the full-spectrum case is shown in Fig. \hyperref[fig:QuasiSat]{12}, alongside a quasilinear saturation estimate. The quasilinear estimate of the heat
flux for a single $k_y$ mode is calculated as \cite{FableQuasiLin,LapQuasiLin},

   \vspace*{-.5cm}
   \begin{equation}
      Q_{k_y}^{QL} = A_0\frac{(\gamma/\langle k_{\perp}^2\rangle)^2}{\vert\phi_{0,k_y}(0)\vert^2}Q_{k_y}^{\textrm{lin.}},
   \end{equation}
with $Q_{k_y}^{\textrm{lin.}}$ representing the linear simulation results for the heat flux and $\phi_{0,k_y}(0)$ the linear electrostatic potential at $k_x = z = 0$.
$A_0$ represents a constant of proportionality, and $\langle k_{\perp}^2\rangle$ is the ballooning-angle-averaged perpendicular wavenumber, defined as \cite{LapQuasiLin},

   \vspace*{-.5cm}
   \begin{equation}
      \langle k_{\perp}^2\rangle = \frac{\sum_{k_x}\int (g^{xx}k_x^2 + 2g^{xy}k_xk_y + g^{yy}k_y^2)\vert\phi_k(z)\vert^2dz}
                                        {\sum_{k_x}\int\vert\phi_k(z)\vert^2dz}.
   \end{equation}
Here, $\phi_k$ represents a Fourier mode of the electrostatic potential perturbation, and $g^{\mu\nu}=\nabla\mu\cdot\nabla\nu$ gives the metric tensor coefficients of the GENE
coordinate system. A sum over all $k_x$ values was required for the drop-off of saturation levels at low $k_y\rho_i$.

\quad This model of mode saturation describes a balance of the unstable growth of the instability with turbulent diffusion based on a mixing-length estimate \cite{JenkoSat1,JenkoSat2,
BourdSat}, and gives notable agreement with the nonlinear heat flux spectrum in the intermediate scale. The disparity between the quasilinear model and the nonlinear heat flux spectrum
is greatest for ETG modes with wavenumbers higher than $k_y\rho_i \approx 12$. This suggests other, stronger saturation mechanisms for these modes. While new effects, such as toroidal
inverse-cascading \cite{LinFluidETG,ChenFluidETG,KimETG}, may play a role in the full-spectrum case, the difference in the heat flux spectra also agrees with the transition in ETG-ZF
dynamics found to occur around $k_y\rho_i = 15$ for the single-mode results. The more abrupt ZF response and quicker saturation of the higher $k_y$ modes may be consistent with secondary
instability theory \cite{DorlandJenkoETG1,DorlandJenkoETG2}, where a $\vert\phi_n\vert \sim \gamma/\langle k_{\perp}^4\rangle$ saturation model for ETG mode amplitudes predicts a steeper
drop-off near $k_y\rho_i \approx 12$ than that of the quasilinear mixing-length estimate.

\quad Finally, we present a comparison of turbulent and neoclassical transport levels at both ion and electron scales. Because the electron-scale case takes the ion
temperature gradient to zero, we can compare the electron-scale thermal diffusivity to that of the ion-scale ITG case with adiabatic electrons to understand the importance
of regulation by zonal flows at each scale. In units normalized to the specific species of interest, the thermal diffusion coefficients due to electrostatic turbulence
are $\langle\chi_{\textrm{ES}}\rangle_i = 0.7 \rho_i^2v_{Ti}/L_{Ti}$ for the ion-scale ITG case and $\langle\chi_{\textrm{ES}}\rangle_e = 2.8 \rho_e^2v_{Te}/L_{Te}$ for
the electron-scale ETG case, where $v_{Ts}$ is the thermal velocity, $\sqrt{2T_s/m_s}$, for a species $s$. This suggests that the ETG-driven zonal flows don't regulate ETG
turbulence as strongly as the isomorphic counterpart ITG turbulence is regulated by ITG-driven zonal flows.

\quad The neoclassical transport values were calculated using GENE for both the ion and electron scale cases. Given in units of $\chi_{gB}$ from Table \hyperref[tableOne]{I},
the neoclassical thermal diffusivites are $\langle\chi_{\textrm{neo}}\rangle_i = 0.14\chi_{\textrm{gB}}$ and $\langle\chi_{\textrm{neo}}\rangle_e = 0.004\chi_{\textrm{gB}}$.
The neoclassical values are in close agreement with the theoretical expectation that $\chi_i = \sqrt{m_i/m_e}\chi_e$, and are negligible compared to the turbulent thermal
diffusivities, $\langle\chi_{\textrm{ES}}\rangle_i = 6.95\chi_{\textrm{gB}}$ and $\langle\chi_{\textrm{ES}}\rangle_e = 0.328\chi_{\textrm{gB}}$. The late-time heat flux
spectrum peaks in the intermediate scale at $k_y\rho_i = 10.6$ with $\langle Q_{ES}\rangle_e = 0.66Q_{\textrm{gB}}$ and drops off to $\langle Q_{ES}\rangle_e = 0.10Q_{gB}$
and $0.11Q_{gB}$ for $k_y\rho_i = 4.24$ and $16.96$ respectively. These values are in good agreement with the theoretical expectation that $Q/Q_{gB} \sim \mathcal{O}(0.01)-
\mathcal{O}(0.1)$ (in the units of Table \hyperref[tableOne]{I}) for the intermediate-scale ETG modes \cite{Chen_ETG_ZF}.

\section{Discussion}

\quad We have shown, using the single-mode nonlinear simulations, that the NLSE model \cite{Chen_ETG_ZF} accurately describes the zonal flow generation mechanism by
intermediate-scale ETG modes and that it provides a theoretical understanding for the slow growth of long-wavelength zonal flows into the long-term quasi-saturated state.
As the NLSE model considers only a single ETG mode for a practicable analysis, one cannot say conclusively that the same is true of the full-spectrum nonlinear results.
However, in the full-spectrum case the high-$k_y$ ETG modes are quickly saturated by a stronger ZF response as compared to the intermediate-scale ETG modes. The
intermediate-scale ETG modes then drive exponential ZF mode growth initially, and slow, algebraic ZF mode growth as they are suppressed in the late stage. This result
is in good agreement with the NLSE model for intermediate-scale ETG-ZF dynamics, as well as various long time, saturated electron-scale \cite{ParkerZF,ColyerETG}
and multiscale \cite{HowardLmode,HollandHmode} flux-tube simulations.

   \begin{figure*}[!ht]
      \centering
      \begin{minipage}[!t]{.5\textwidth}
         \vspace*{2.7cm}
         \label{fig:HeatFluxConv}
         \includegraphics[width=.9\linewidth]{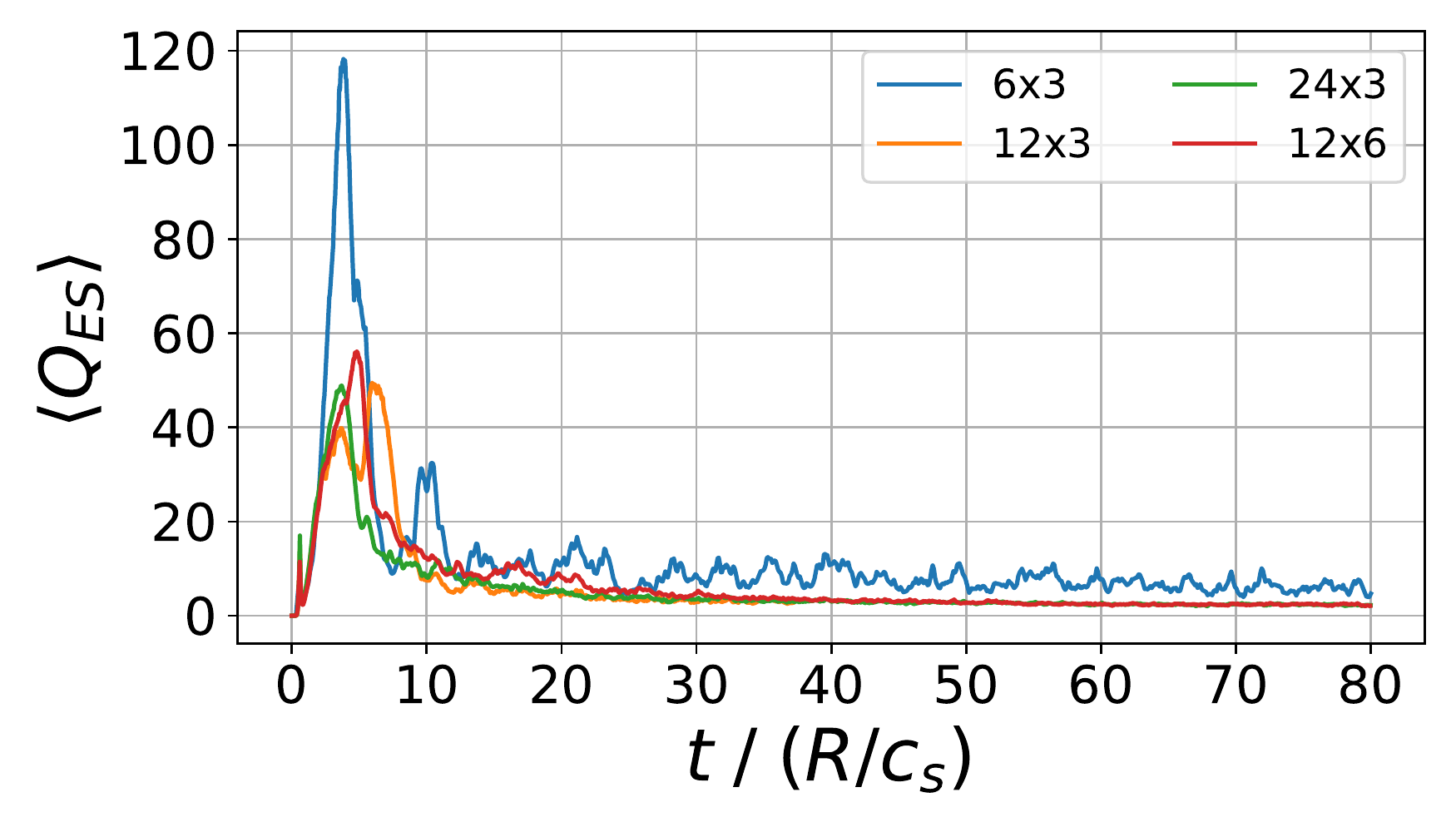}
         \caption{Convergence of the electrostatic electron heat flux for the various box size cases compared to the original $6\rho_i\times 3\rho_i$ case.}
      \end{minipage}%
      \hspace*{.5cm}
      \begin{minipage}[!t]{.5\textwidth}
         \centering
         \includegraphics[width=.9\linewidth]{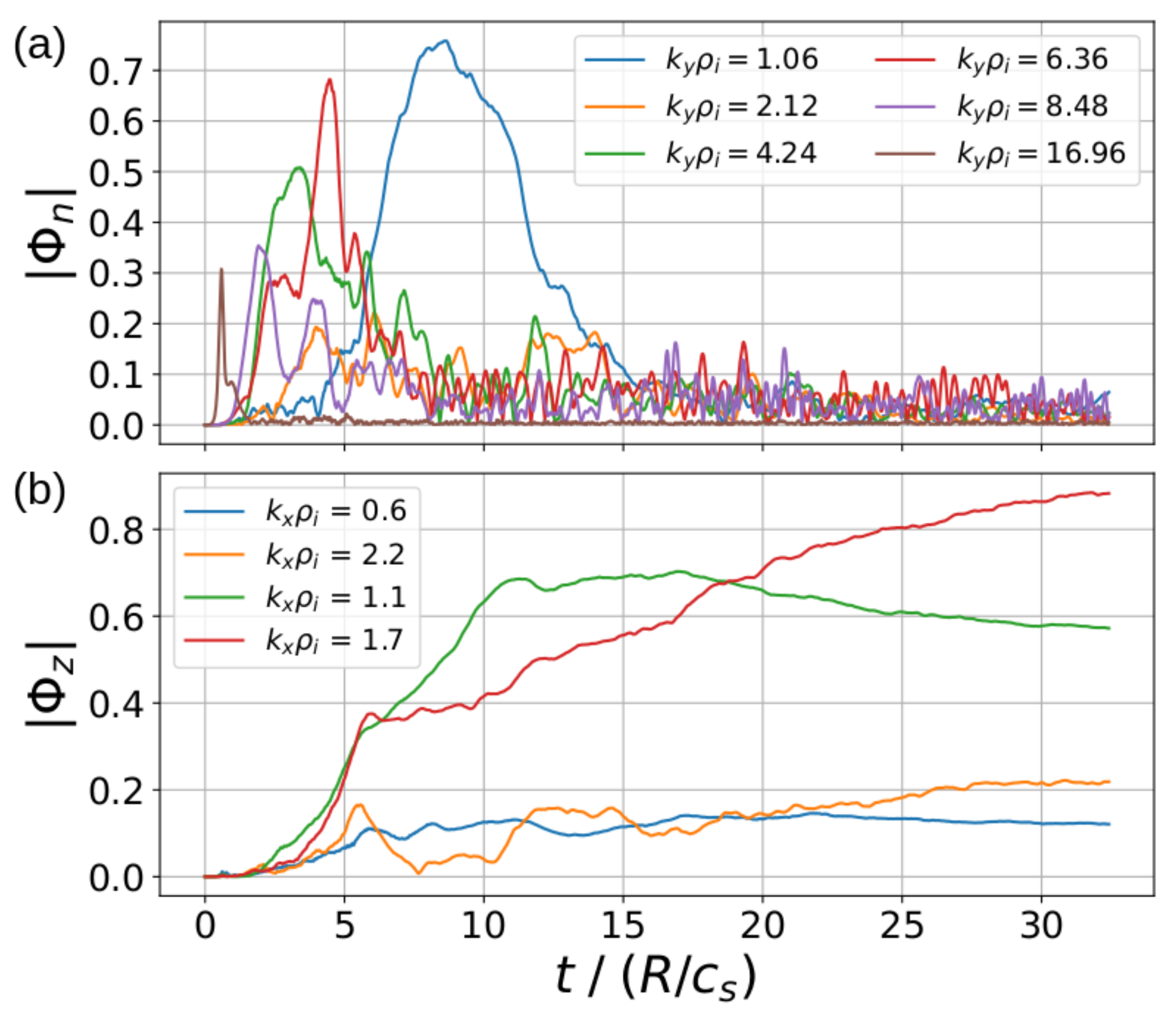}
         \caption{(a) ETG and (b) four strongest final ZF mode time traces for the $12\rho_i\times 6\rho_i$, large-y box size. All data is averaged over $z$.}
         \label{fig:ConvETG}
      \end{minipage}
   \end{figure*}
The final transport levels for the full-spectrum case are in similar ranges found in thorough electron-scale CBC benchmarks which compare well with experimental observations
\cite{Nevins06}. As the zonal flows are driven at long electron-scale wavelengths, multiscale effects could become important and ETG-driven zonal flows may have an effect
on ion-scale turbulence. This effect where intermediate-scale zonal flows contribute to ion-scale turbulence suppression has been reported in large multiscale simulations
\cite{HowardLmode,HollandHmode}.

\section*{Acknowledgements}

\quad This research was supported by the SciDAC-4 project High-fidelity Boundary Plasma Simulation funded by the U.S. Department of Energy (U.S. DOE) Office of Science under
grant DE-SC-000801. Research was also carried out under U.S. DOE  grant DE-FG02-08ER54954.

\appendix

\section{\label{sec:ConvergenceNL}Nonlinear Convergence Tests}

\quad This section details a ``full-spectrum" nonlinear simulation box-size convergence study. The perpendicular box size is varied in terms of the basic $6\rho_i\times3\rho_i$
electron-scale box size shown in Fig. \hyperref[fig:BaseETG]{2}. The collisionality was set to the reference value discussed in Sec. \ref{sec:FullSpec}. Four perpendicular domain
sizes are presented: $6\rho_i\times3\rho_i$, $12\rho_i\times3\rho_i$, $24\rho_i\times3\rho_i$, and $12\rho_i\times6\rho_i$. These cases consider the importance of correctly
resolving the ETG streamer lengths and the longest wavelength zonal flow modes. Additionally, the inclusion of longer ETG mode wavelengths is considered in the $12\rho_i\times6\rho_i$
case to verify the findings of Section \ref{sec:SingleModeResults}. The number of radial gridpoints was increased in each simulation to retain the original resolution.

\quad The electron heat flux for each case is shown over time in Fig. \hyperref[fig:HeatFluxConv]{13}. One can clearly see that the increase in radial dimension is necessary
to correctly resolve the heat flux. The late-stage zonal flows of the $6\rho_i\times3\rho_i$ case, as shown in Fig. \hyperref[fig:LateFourierPhi]{3(c)}, are peaking at the
longest mode allowable and the box size must be increased to correctly resolve the longest modes. Allowing for longer wavelength ZF modes leads to stronger regulation of the
heat flux as seen in Fig. \hyperref[fig:HeatFluxConv]{13}.

\quad The time evolution of the ETG modes for the $12\rho_i\times6\rho_i$ case is shown in Fig. \hyperref[fig:ConvETG]{14(a)}, and the time evolution of the four strongest ZF modes
at the final time is shown in Fig. \hyperref[fig:ConvETG]{14(b)}. These results are qualitatively similar to the $24\rho_i\times 3\rho_i$ case shown in Figs. \hyperref[fig:CollZonal]{10(b)}
and \hyperref[fig:CollETG]{11}. In this new case, the longest wavelength ETG mode, $k_y\rho_i=1.06$, grows to the highest level. However, it can be seen in Fig. \hyperref[fig:ConvETG]{14(b)}
that from non-dimensional times 5-10 $t/(R/c_s)$, when the longest ETG mode is dominant, the zonal flows are already in the final, slowly growing stage. This result indicates that the
strongest ZF modes are largely being affected by the intermediate-scale ETG modes, not the longest wavelength ETG mode, and confirms the results found in Sec. \ref{sec:SingleModeResults}
which showed little zonal flow generation outside the intermediate-scale range. As increasing $L_y$ from the original $3\rho_i$ size had no effect on the final quasi-saturated state,
the largest $L_x$ case considered, $24\rho_i\times3\rho_i$, was chosen for the full-spectrum investigation discussed in Sec. \ref{sec:FullSpec}.

\bibliographystyle{unsrtnat} 

\end{document}